\documentclass[a4paper,twoside,11pt,twocolumn]{article}
\usepackage{geometry}
\usepackage{sectsty} 

\usepackage[english]{babel} 
\usepackage[english=american]{csquotes}
\usepackage{amsmath,amsfonts,amsthm} 
\usepackage[svgnames]{xcolor}
\usepackage{subcaption, graphicx, booktabs}
\usepackage[separate-uncertainty = true,multi-part-units=single]{siunitx}
\usepackage{hyphenat}
\usepackage{stfloats}
\usepackage{comment}
\usepackage{lipsum}
\usepackage{acronym}
\usepackage{chemformula}
\usepackage{parskip}
\usepackage[backend=biber,style=vancouver,sorting=none, url=false]{biblatex}
\usepackage{authblk}
\usepackage[hidelinks]{hyperref}

\usepackage{setspace}

\allsectionsfont{\usefont{OT1}{phv}{b}{n}}
\graphicspath{{Figures/}}
\DeclareCaptionLabelFormat{subcapformat}{\textbf{#2}}
\DeclareSIUnit\angstrom{\protect \text  {Å}}
\title{Structure-property relations of silicon oxycarbides studied using a machine learning interatomic potential}

\author[1,*]{Niklas Leimeroth}
\author[1]{Jochen Rohrer}
\author[1]{Karsten Albe}
\affil[1]{Institut für Materialwissenschaft, Technische Universität Darmstadt, Otto-Berndt-Strasse 3, 64287 Darmstadt, Germany}
\affil[*]{Corresponding author email: leimeroth@mm.tu-darmstadt.de}

\date{\today}

\newcommand{\PACEMAKER}{{\texttt{\detokenize{pacemaker}}}}
\newcommand{\LAMMPS}{{\texttt{\detokenize{LAMMPS}}}}
\newcommand{\KOKKOS}{{\texttt{\detokenize{KOKKOS}}}}
\newcommand{\MLIP}{{\texttt{\detokenize{MLIP}}}}
\newcommand{\ASE}{{\texttt{\detokenize{ASE}}}}
\newcommand{\PACKMOL}{{\texttt{\detokenize{PACKMOL}}}}

\newcommand{\stoicho}{Si$_{0.4}$O$_{0.4}$C$_{0.2}$}
\newcommand{\pmsq}{Si$_{0.25}$O$_{0.5}$C$_{0.25}$}
\newcommand{\rdtwo}{Si$_{0.25}$O$_{0.25}$C$_{0.5}$}
\newcommand{\silres}{Si$_{0.125}$O$_{0.125}$C$_{0.75}$}
\newcommand{\rdsix}{Si$_{0.121}$O$_{0.121}$C$_{0.758}$}

\begin{document}

\acrodef{DFT}[DFT]{density-functional theory}
\acrodef{MD}[MD]{molecular dynamics}
\acrodef{ACE}[ACE]{Atomic Cluster Expansion}
\acrodef{MTP}[MTP]{Moment Tensor potential}
\acrodef{MLIP}[MLIP]{machine learning interatomic potential}
\acrodef{IP}[IP]{Interatomic potential}
\acrodef{SOAP}[SOAP]{Smooth Overlap of Atomic Positions}
\acrodef{NNP}{Neural Network potential}
\acrodef{SiOC}[Si-O-C]{silicon oxycarbide}
\acrodef{AL}[AL]{active learning}
\acrodef{BF}[BF]{bulk fragment}
\acrodef{Ats}[Ats]{isolated atoms}
\acrodef{GrAts}[GrAts]{graphite flakes and isolated atoms}
\acrodef{PD}[PD]{polymer derived}
\acrodef{PMSQ}[PMSQ]{Polymethylsilsesquioxane} 
\acrodef{E}[E]{Young's modulus}

\maketitle

\begin{abstract}
  Silicon oxycarbides show outstanding versatility due to their highly tunable composition and microstructure.
  Consequently, a key challenge is a thorough knowledge
  of structure-property relations in the system.
  In this work, we fit an atomic cluster expansion potential
  to a set of actively learned DFT training data spanning a wide configurational space.
  We demonstrate the ability of the potential
  to produce realistic amorphous structures and
  rationalize the formation of different morphologies of the turbostratic free carbon phase.
  Finally, we relate the materials stiffness to its composition and microstructure,
  finding a delicate dependence on Si-C bonds
  that contradicts commonly assumed relations to the free carbon phase.
\end{abstract}

\section{Introduction}

\Acp{SiOC} are highly versatile materials that
combine remarkable structural and functional properties.
Among them are high temperature resistance,
good mechanical strength \cite{liAdditiveManufacturingLightweight2020,barriosReviewEvolutionNanostructure2020}, great creep- and corrosion-resistance \cite{papendorfHighTemperatureCreepBehavior2013,stablerHightemperatureCreepBehavior2016},
as well as piezoresistivity and the ability to reversibly store Li$^+$, Na$^+$ and K$^+$ \cite{melziderilEffectUltrafastPyrolysis2023,vallachirawarriamsasikumarStructuralDesignPolymerDerived2016,
  stablerSiliconOxycarbideGlasses2018,chandraUnderstandingLithiumSodium2022,}.
These properties make them interesting for applications in very different fields,
such as protective coatings \cite{luPolymerDerivedSilicon2018},
energy storage \cite{wenSibasedPolymerderivedCeramics2022}
and biomedicine
\cite{arango-ospinaReviewSiliconOxycarbide2020}.

Despite the plethora of desirable properties and
intensive research, open questions
about structural features and their
relation even to basic characteristics such as the \ac{E} remain.
From NMR measurements it is known that \acp{SiOC} consist of corner-shared SiO$_x$C$_{4-x}$ tetrahedra
with carbidic sp$^3$-hybridized carbon
\cite{widgeon29Si13CSolidState2010,meraPolymerDerivedSiCN2013}
and a segregated secondary phase with sp$^2$-hybridized turbostratic carbon
\cite{soraruStructuralCharacterizationHighTemperature1995,rothUVRamanSpectroscopy2016}.
The detailed nanostructure, however, remains elusive and
sensitively depends on the composition, precursor structure and processing conditions.
For example, it is still unclear in which form the turbostratic carbon
is present in \ac{SiOC}.
Scarmi et al. \cite{scarmiRoleCarbonUnexpected2005} and  Saha et al. \cite{sahaModelNanodomainsPolymerDerived2006}
argued that interpenetrating networks of graphene-like carbon and silica-rich mixed tetrahedra domains are formed,
based on the high creep resistance of the material.
In contrast, Widgeon et al. \cite{widgeon29Si13CSolidState2010}
found that a model with graphitic inclusions embedded
in a silica rich matrix result in a better match of the mass fractal dimensions
of mixed tetrahedra.

Here, atomistic simulations may help in understanding structure formation
and structure-property relations of \ac{SiOC} compounds at the nanoscale.
As shown for example in a series of studies by Kroll \cite{krollModellingSimulationAmorphous2003,krollModelingFreeCarbon2005,krollSearchingInsightAtomistic2010},
ab-initio \ac{MD} simulations can be employed to investigate
structural details, energetics and elastic properties of \ac{SiOC},
but are limited to structures consisting of a few hundred atoms and simulation times
on the order of tens of picoseconds.
Consequently, the heterogeneity on the nanoscale intrinsic to \ac{SiOC} can not be reproduced in full extend.
Large-scale \ac{MD} simulations, on the other hand, require suitable interatomic potentials
and the complex nature of the strongly directional
covalent bonds is hard to capture in empirical formulas
like bond order potentials.
Recent studies showed that the ReaxFF framework
\cite{vanduinReaxFFReactiveForce2001} allows
to study specific aspects of the Si-O-C system.
For example, Newsome et al. simulated the oxidation of silicon carbide \cite{newsomeOxidationSiliconCarbide2012}.
Soria et al. investigated organic molecules on silicon surfaces \cite{soriaSiReaxFFReactive2018},
Gao et al. and Ponomarev et al. simulated the pyrolysis of specific polymers to amorphous \ac{SiOC} \cite{gaoReactiveDynamicsSimulation2018, ponomarevReactiveForceField2019}.
These potentials, however, have a limited scope
and require a reparametrization for each application.
Furthermore, only two out of the four parameter sets are publicly available \cite{newsomeOxidationSiliconCarbide2012,soriaSiReaxFFReactive2018}.
Here, modern \acp{MLIP} offer an alternative approach to describe complex systems
over a wide compositional and structural range at similar computational cost,
but at the expanse of requiring more training data.
Recent studies have shown the successful application of \acp{MLIP} to
carbon \cite{
  khaliullinGraphitediamondPhaseCoexistence2010,
  deringerMachineLearningBased2017,
  wenHybridNeuralNetwork2019,
  caroOptimizingManybodyAtomic2019,
  roweAccurateTransferableMachine2020,
  shaiduSystematicApproachGenerating2021,
  wangDeepLearningInteratomic2022,
  qamarAtomicClusterExpansion2023,
},
silicon \cite{
  bartokMachineLearningGeneralPurpose2018,
  qianThermalConductivityModeling2019,
  yokoiNeuralnetworkInteratomicPotential2020,
  georgeCombiningPhononAccuracy2020,
  hamedaniInsightsPrimaryRadiation2020,
  lysogorskiyPerformantImplementationAtomic2021,
}, \ch{SiO2} \cite{
  novikovImprovingAccuracyInteratomic2019,
  balyakinDeepMachineLearning2020,
  kobayashiSelflearningHybridMonte2021,
  erhardMachinelearnedInteratomicPotential2022,
}
and SiC \cite{kutzhanovSiCNanocompositesEnhanced2021,liuDeepLearningInteratomic2023}.
In this work, we present an \ac{ACE} potential for the \ac{SiOC} system fitted
to an extensive database generated using the \ac{AL} capabilities of
\acp{MTP} \cite{novikovMLIPPackageMoment2021} and \ac{ACE} \cite{lysogorskiyActiveLearningStrategies2023}.
We show that the potential
achieves a high accuracy for a wide compositional range
and that it can be used to produce realistic amorphous \ac{SiOC} structures.
We investigate structural features, formation energies and Young's moduli for samples with varying compositions
and precursor configurations.
Thereby, we find that the structure model containing graphene like sheets and the
graphitic inclusions in a silica matrix
are both likely to describe the structure of \ac{SiOC}, but occur at different stages of processing.
Furthermore, we establish relations of \acl{E} to the fraction of Si-C bonds in mixed SiO$_x$C$_{4-x}$ tetrahedra
and the silica volume.

\section{Methods}

\begin{figure*}[tb!]
  \centering
  \begin{subfigure}[t]{0.63\linewidth}
    \centering
    \includegraphics[width=0.95\linewidth]{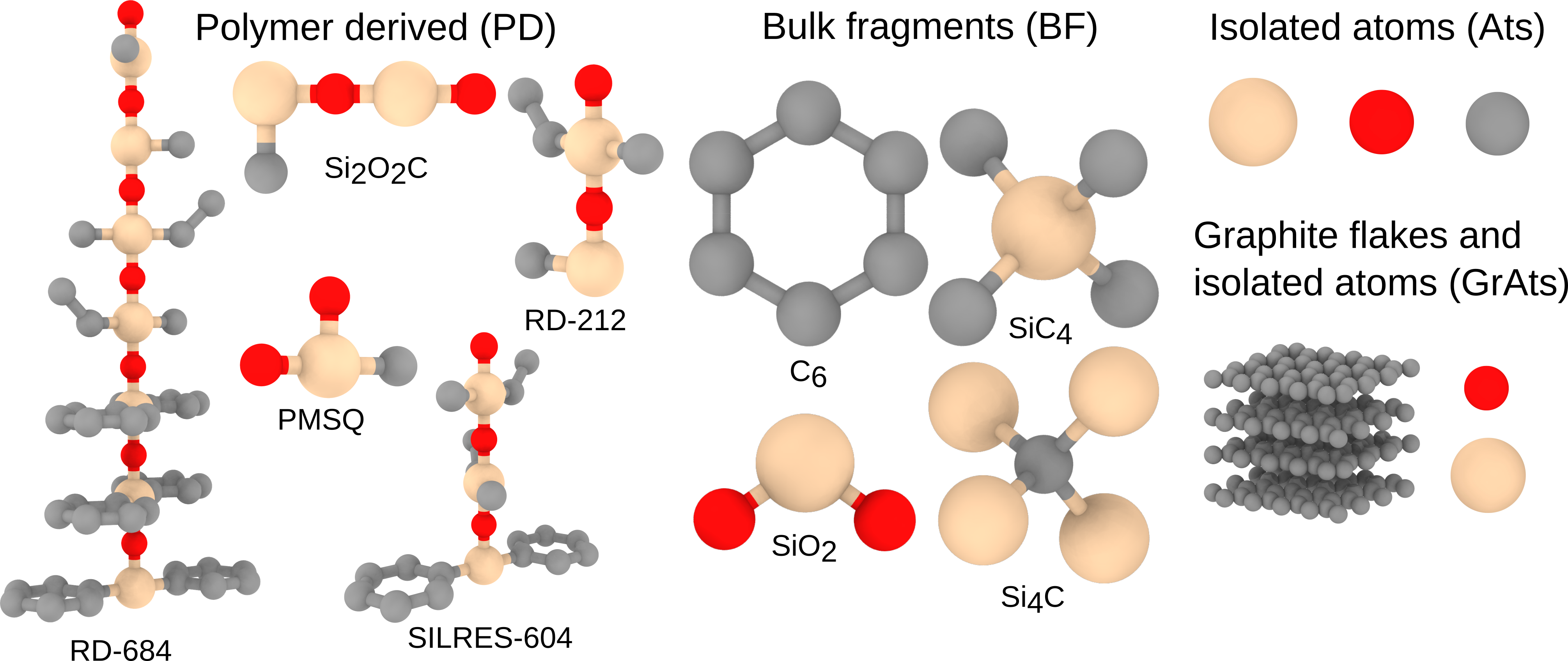}
    \caption{}
    \label{fig:BuildingBlock}
  \end{subfigure}%
  \begin{subfigure}[t]{0.33\linewidth}
    \centering
    \includegraphics[width=0.95\linewidth]{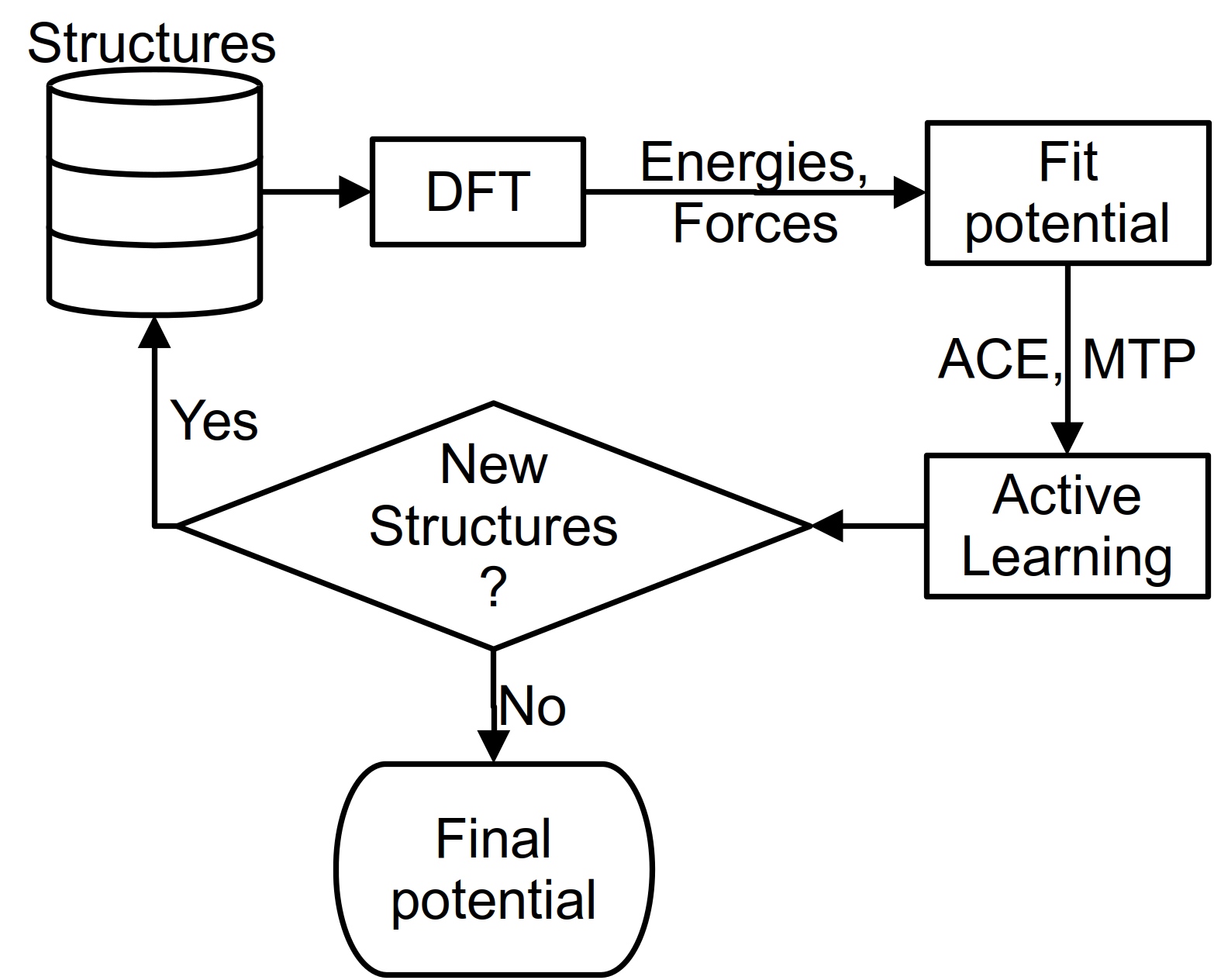}
    \caption{}
    \label{fig:ActiveLearning}
  \end{subfigure}
  \begin{subfigure}{0.7\linewidth}
    \centering
    \includegraphics[width=0.8\linewidth]{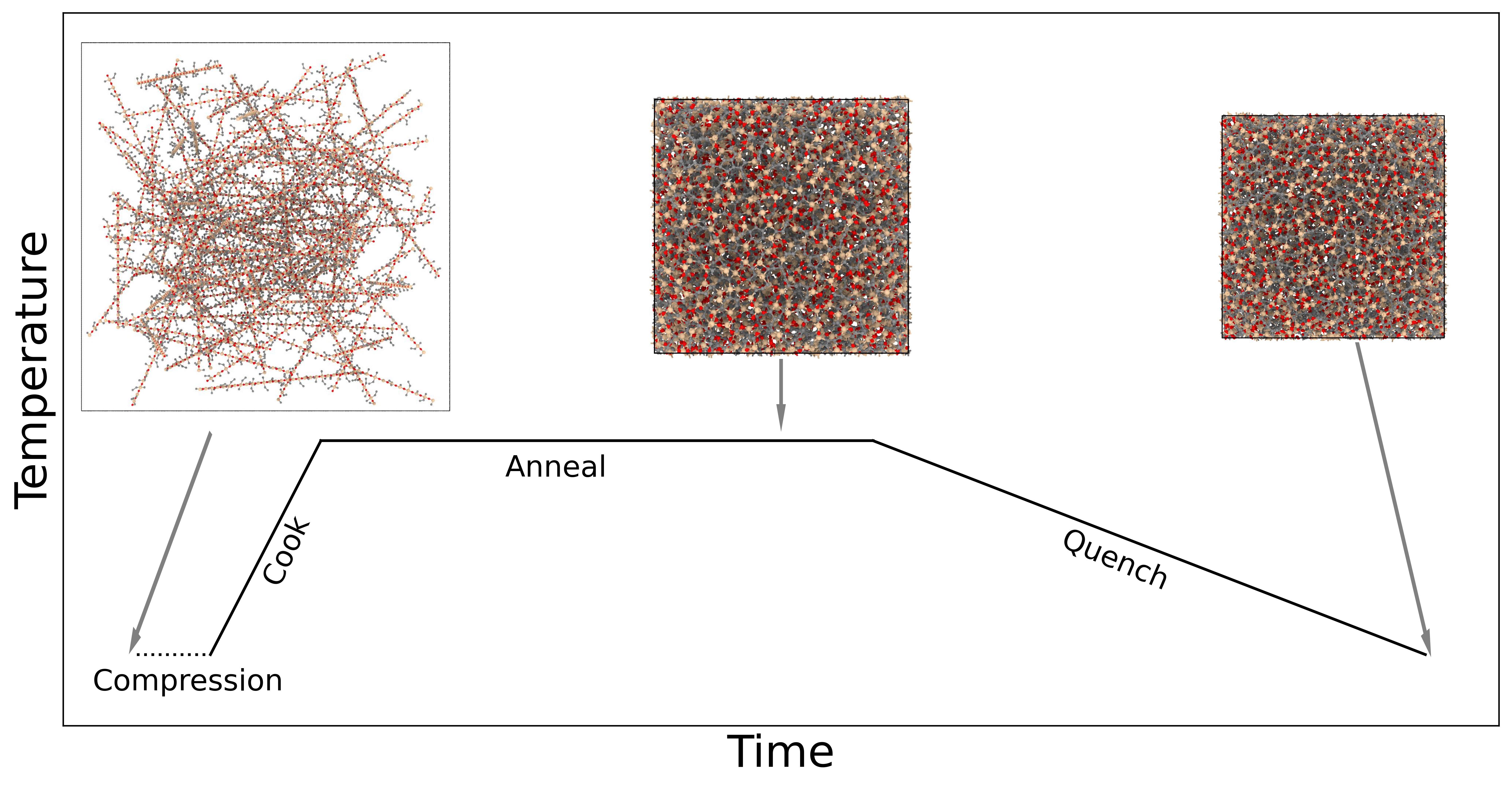}
    \caption{}
    \label{fig:SchematicCookAndQuench}
  \end{subfigure}
  \caption{
    \textbf{Si-O-C structure generation and active-learning strategy.}
    Most initial structures for the \ac{AL} process were generated by densely packing smaller building blocks using the \PACKMOL{} program.
    These building blocks and the nomenclature used to describe them throughout this work are shown in (\subref{fig:BuildingBlock}).
    Si is shown in beige, O in red and C in gray.
    PD structures are stripped of H atoms.
    GrAts building blocks were used to produce large sample structures, but not in the \ac{AL} process.
    The \ac{AL} procedure employed to iteratively improve the training data is schematically shown in (\subref{fig:ActiveLearning}).
    Structures for \ac{AL} and samples used in the analysis were produced using a cook and quench simulations depicted in (\subref{fig:SchematicCookAndQuench}).
    The compressions step shown as dotted line was only applied for PD and GrAts structures,
    where it was necessary to obtain nonporous bulk samples.
    For structures made from BFs or Ats it was not necessary,
    because the initial packing already leads to reasonable densities.
  }
  \label{fig:TrainingDataAL}
\end{figure*}
\begin{table}[btp!]
  \centering
  \begin{tabular}{@{}llll@{}}
    \toprule
                 & X$_{\mathrm{Si}}$ & X$_{\mathrm{O}}$ & X$_{\mathrm{C}}$ \\ \midrule
    Si$_2$O$_2$C & 0.4               & 0.4              & 0.2              \\
    PMSQ         & 0.25              & 0.5              & 0.25             \\
    RD-212       & 0.25              & 0.25             & 0.5              \\
    SILRES-604   & 0.125             & 0.125            & 0.75             \\
    RD-684       & $\approx$0.121    & $\approx$0.121   & $\approx$0.758   \\ \bottomrule
  \end{tabular}%
  \caption{Compositions of hydrogen stripped polymers, sorted from low to high carbon content.
    These compositions were used to generate \acf{PD}, \acf{BF}, \acf{Ats} and \acf{GrAts} samples (See Fig. \ref{fig:BuildingBlock}).}
  \label{tab:Compositions}
\end{table}

\subsection{Training and testing data}

The training data for the potential was produced as follows.
Initial Si, O and Si-O structures were taken from an \ac{ACE}
potential previously fitted for the Si-O system \cite{erhardModellingAtomicNanoscale2023}.
Pure C structures were generated using the \ASE{} package
\cite{hjorthlarsenAtomicSimulationEnvironment2017}
and SiC structures were taken from the materials project database
\cite{jainCommentaryMaterialsProject2013}.
Si-O-C structures with varying compositions were generated with two different procedures.
Firstly, structures were produced to match the expected coordination,
i.e. 4-fold for Si and C and 2-fold for O \cite{rohrerSiSncontainingSiOCNbased2017,sicSiCOCeramicsStorage}.
Additional structures were then generated using an \ac{AL} procedure,
schematically shown in Fig. \ref{fig:TrainingDataAL}.
We employed \PACKMOL{} \cite{martinezPACKMOLPackageBuilding2009},
which implements an algorithm to densely pack structural fragments
while keeping adjustable minimal distances between atoms \cite{martinezPackingOptimizationAutomated2003}.
With this approach
structures based on polymers inspired from \ac{PMSQ},
the polyorganosiloxanes RD-212, RD-684 and SILRES-604
and a fully artificial polymer made from \ch{Si2O2C} monomers were generated
with removed H atoms.
In the following we will use the compositions,
instead of polymer names as shown in Tab. \ref{tab:Compositions}.
Stripping H from the polymers avoids a massive extension of the necessary training data,
as the configuration space for a quarternary system is much larger than for a ternary,
but keeps the polymeric backbone.
We expect that this has a negligible influence for
final structures produced with pyrolysis temperatures of \SIrange{1273}{1523}{K} and higher,
because H evaporates during synthesis and the remaining amount of H atoms in them is very small \cite{wenFateRoleSitu2020a,delverdierThermalBehaviorPolymerderived1993,pantanoSiliconOxycarbideGlasses1999}.

In order to cover a wider variety of compositions and structural features, we also produced configurations made
from \acf{Ats} and \acfp{BF} \ch{SiO2}, \ch{SiC4}, \ch{CSi4} and \ch{C6}
as shown in Fig. \ref{fig:BuildingBlock}.
For the structures generated directly from \ac{Ats} and \acp{BF} the composition was varied,
with Si and C concentrations ranging from \SIrange{0}{100}{\%}
and the oxygen concentration from \numrange{0}{2.2} times the Si concentration.
The \ac{AL} capabilities
of \acp{MTP} \cite{podryabinkinActiveLearningLinearly2017} and
\ac{ACE} potentials \cite{lysogorskiyActiveLearningStrategies2023}
were employed to iteratively generate new Si-O-C configurations
as shown in Fig. \ref{fig:ActiveLearning}.
Here, we started with \acp{MTP} because the \ac{AL} capabilities for
\ac{ACE} were implemented only recently \cite{lysogorskiyActiveLearningStrategies2023}.
Furthermore, we  covered a wider variety of structures by using both codes.
The training data set was considered complete when no new structures,
that had a maximum \ac{DFT} force of less than \SI{150}{eV/\angstrom}, were discovered
in cook and quench simulations with temperatures up to \SI{3000}{K} and pressures up to \SI{200}{GPa}.

For the test data set the test data of the Si-O potential \cite{erhardModellingAtomicNanoscale2023} was supplemented with
a separately created set of Si-O-C structures.
These Si-O-C test structures were created with \PACKMOL{}
at varying densities and compositions.
Consequently, the atoms were randomly displaced.

Finally, the training and test data sets were filtered
to not contain structures with a maximum force of more than \SI{150}{eV/\angstrom},
a minimal distance between atoms smaller than \SI{0.6}{\angstrom} or greater than \SI{4}{\angstrom}
or an energy of more than \SI{20}{eV/atom} above the convex hull.

\subsection{Details of potentials}

For the potentials we used a cutoff of \SI{5}{\angstrom}.
The \ac{ACE} and \acp{MTP} were fitted using the
\PACEMAKER{} \cite{lysogorskiyPerformantImplementationAtomic2021,bochkarevEfficientParametrizationAtomic2022}
and \MLIP{} packages \cite{novikovMLIPPackageMoment2021}.
For the \ac{AL} process with \acp{MTP} a level 26 potential was employed.
The intermediate ACE potentials for \ac{AL} were fitted with the
triple embedding $\rho_1^{0.5} + \rho_2^{1} + \rho_3^{2}$ and a total of 2325 basis functions.
In principle, the most straightforward way of increasing the accuracy of \ac{ACE} potentials is to increase the number of basis functions.
However, this considerably increases the computational cost \cite{bochkarevEfficientParametrizationAtomic2022}.
Instead, we tested different embedding terms for the final potential,
which is computationally very cheap.
Here, we found a highly nonlinear sum of expansions $\sum_i^n \phi_i^{\alpha_i}$ with $n=10$ and exponents $\alpha_i$ 0.125, 0.25, 0.375, 0.5, 0.75, 0.875, 1, 1.25, 1.5 and 2
to result in the best testing errors.

\subsection{SiOC sample structures \label{sec:SampleStructures}}

The structure samples used to analyze formation energies,
structural features and elastic properties were produced in a cook and quench
process, as schematically shown in Fig. \ref{fig:SchematicCookAndQuench}.
Here we tested two different degrees of freedom,
the influence of the composition and the effect of the precursor
on the structure and properties of the final sample.
The compositions of the structures correspond to the 5 polymeric compositions shown in Tab. \ref{tab:Compositions}.
As precursors, we employed 4 different types of building blocks shown in Fig. \ref{fig:BuildingBlock}.
The graphite sheets in \ac{GrAts} structures consist of 160 atom 2-layer graphite.
The resulting 20 structures were used in cook and quench simulations with
annealing temperatures of \num{1000}, \num{1500} and \SI{2000}{K}
to obtain a total of 60 sample structures.
Here, the annealing time was \SI{1}{ns}.
As shown in the supplemental material (section S1) this
time is sufficient to reach a steady state regarding different structural features.
The employed quench rate was \SI{1e12}{K/s}.
\ac{PD} and \ac{GrAts} based structures required an additional compression step
before the cook and quench process to obtain nonporous initial structures.
For this purpose, they were equilibrated at \SI{500}{K} with an applied isotropic pressure of \SI{10}{GPa}
for \SI{10}{ps} before heating them up to the annealing temperature.

\subsection{Simulations}

\ac{DFT} calculations were carried out with the same settings as used for a silica
potential previously fitted \cite{erhardMachinelearnedInteratomicPotential2022}
to keep the training data consistent.
The plane wave code VASP
\cite{kresseTheoryCrystalStructures1994,kresseEfficiencyAbinitioTotal1996,kresseEfficientIterativeSchemes1996}
with projector-augmented wave
\cite{kresseUltrasoftPseudopotentialsProjector1999}
pseudopotentials and the SCAN
\cite{sunStronglyConstrainedAppropriately2015}
meta-GGA exchange-correlation were employed with
a plane-wave cutoff of \SI{900}{eV} and a k-spacing of \SI{0.23}{\angstrom^{-1}}.
Classical \ac{MD} simulations were carried out with \LAMMPS{} \cite{thompsonLAMMPSFlexibleSimulation2022},
applying GPU accelerated \KOKKOS{} versions where possible.
If not otherwise noted an NPT ensemble with isotropic \SI{0}{Pa} pressure,
Nosé-Hoover thermo- and barostats and a timestep of \SI{1}{fs} were employed in the simulations.

\begin{figure*}[tb!]
  \centering
  \begin{subfigure}[t]{0.33\linewidth}
    \centering
    \includegraphics[width=0.99\linewidth]{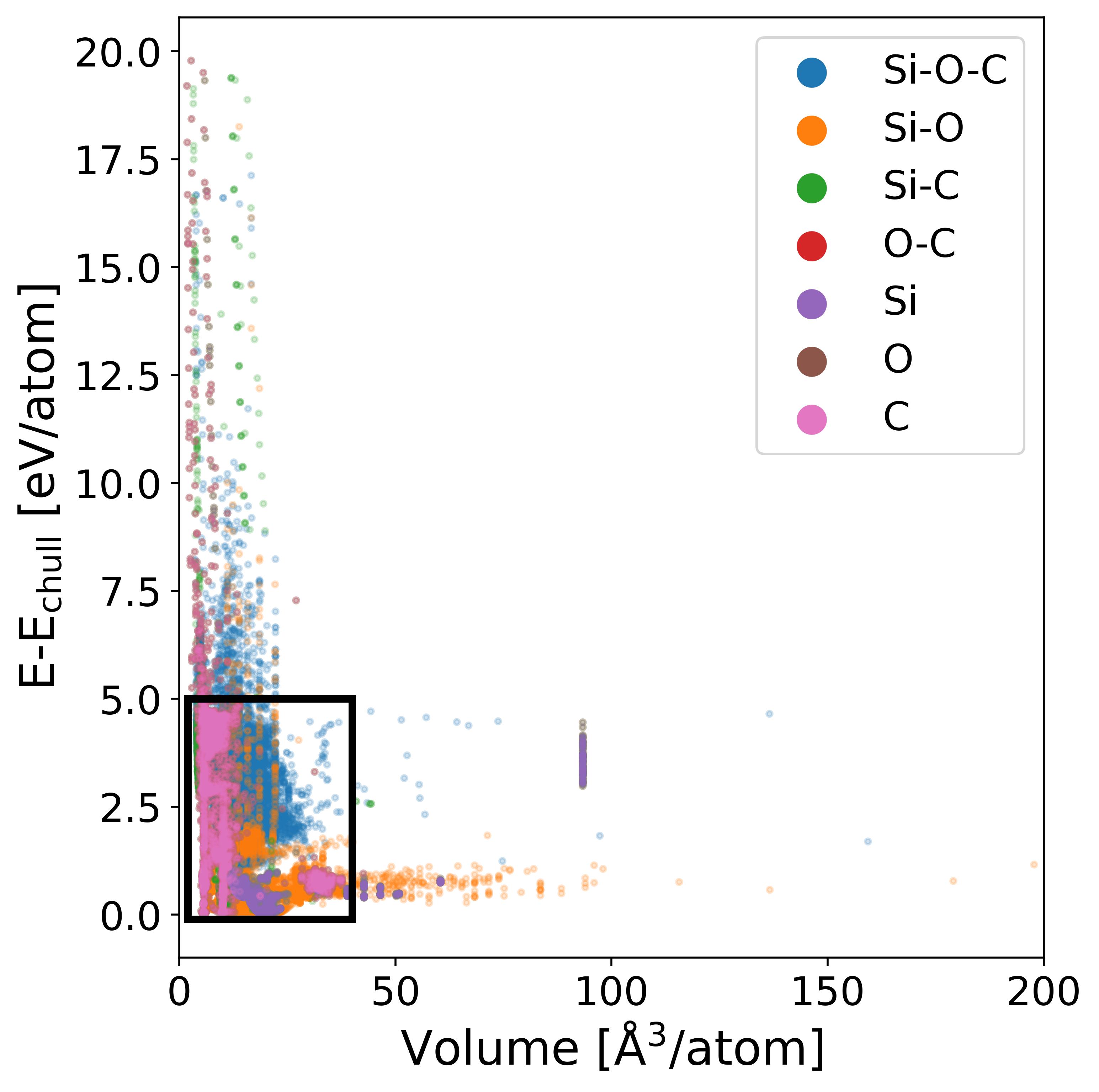}
    \caption{}
    \label{fig:EnergyVol}
  \end{subfigure}%
  \begin{subfigure}[t]{0.33\linewidth}
    \centering
    \includegraphics[width=0.99\linewidth]{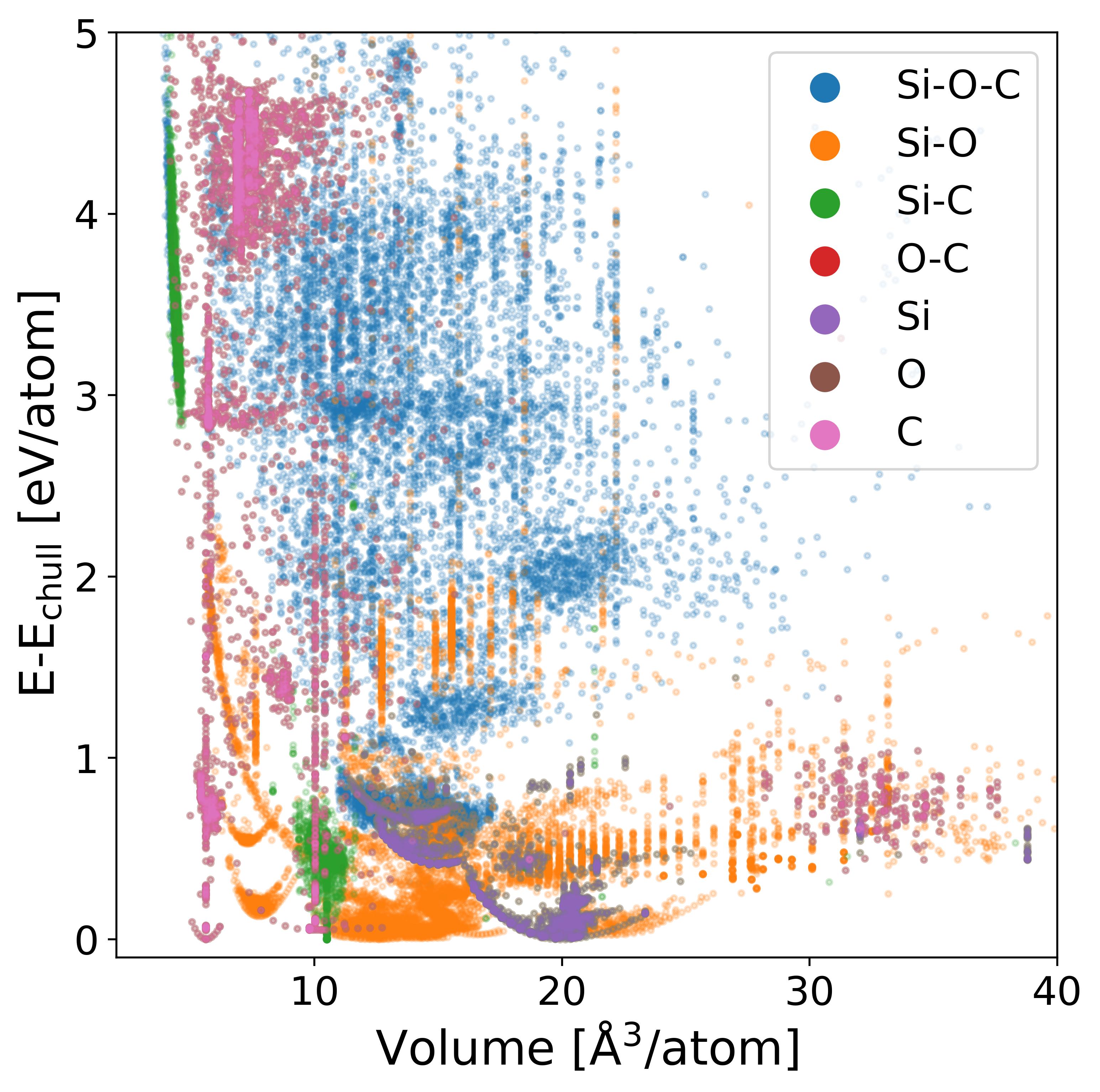}
    \caption{}
    \label{fig:EnergyVolZoomed}
  \end{subfigure}%
  \begin{subfigure}[t]{0.33\linewidth}
    \centering
    \includegraphics[width=0.99\linewidth]{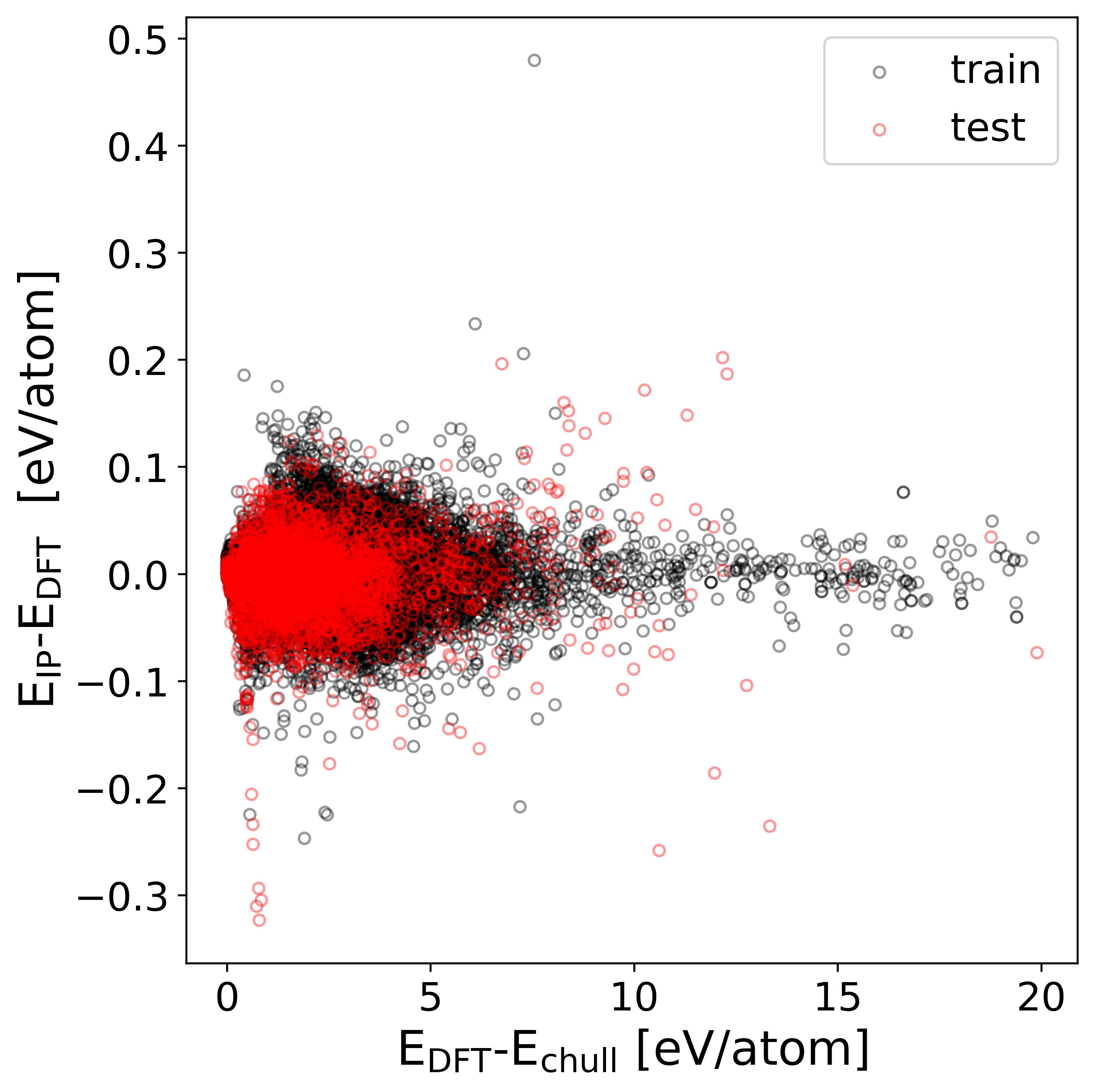}
    \caption{}
    \label{fig:ScatterEnergy}
  \end{subfigure}
  \begin{subfigure}[t]{0.33\linewidth}
    \centering
    \includegraphics[width=0.99\linewidth]{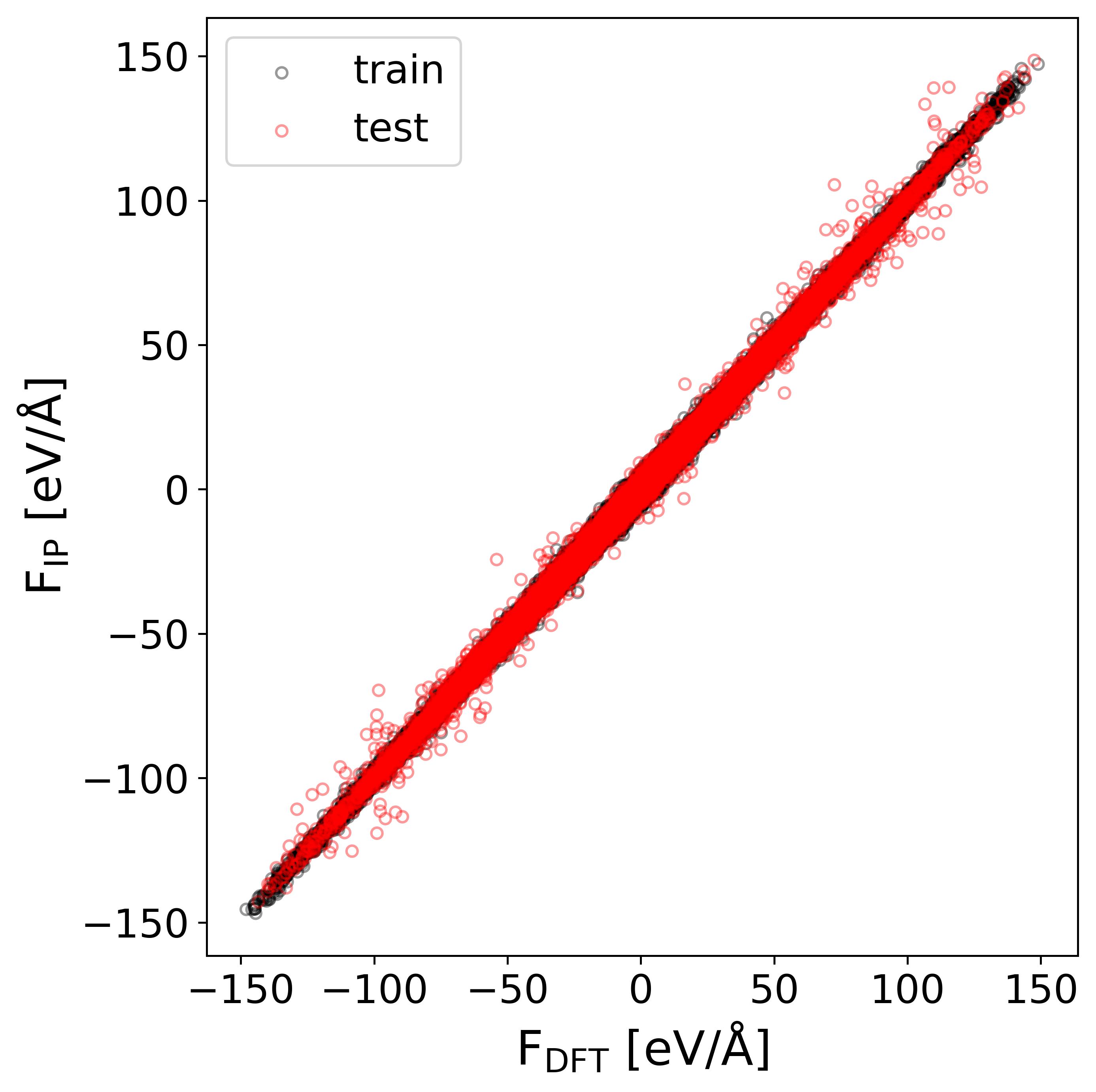}
    \caption{}
    \label{fig:ScatterForce}
  \end{subfigure}%
  \begin{subfigure}[t]{0.33\linewidth}
    \centering
    \includegraphics[width=0.99\linewidth]{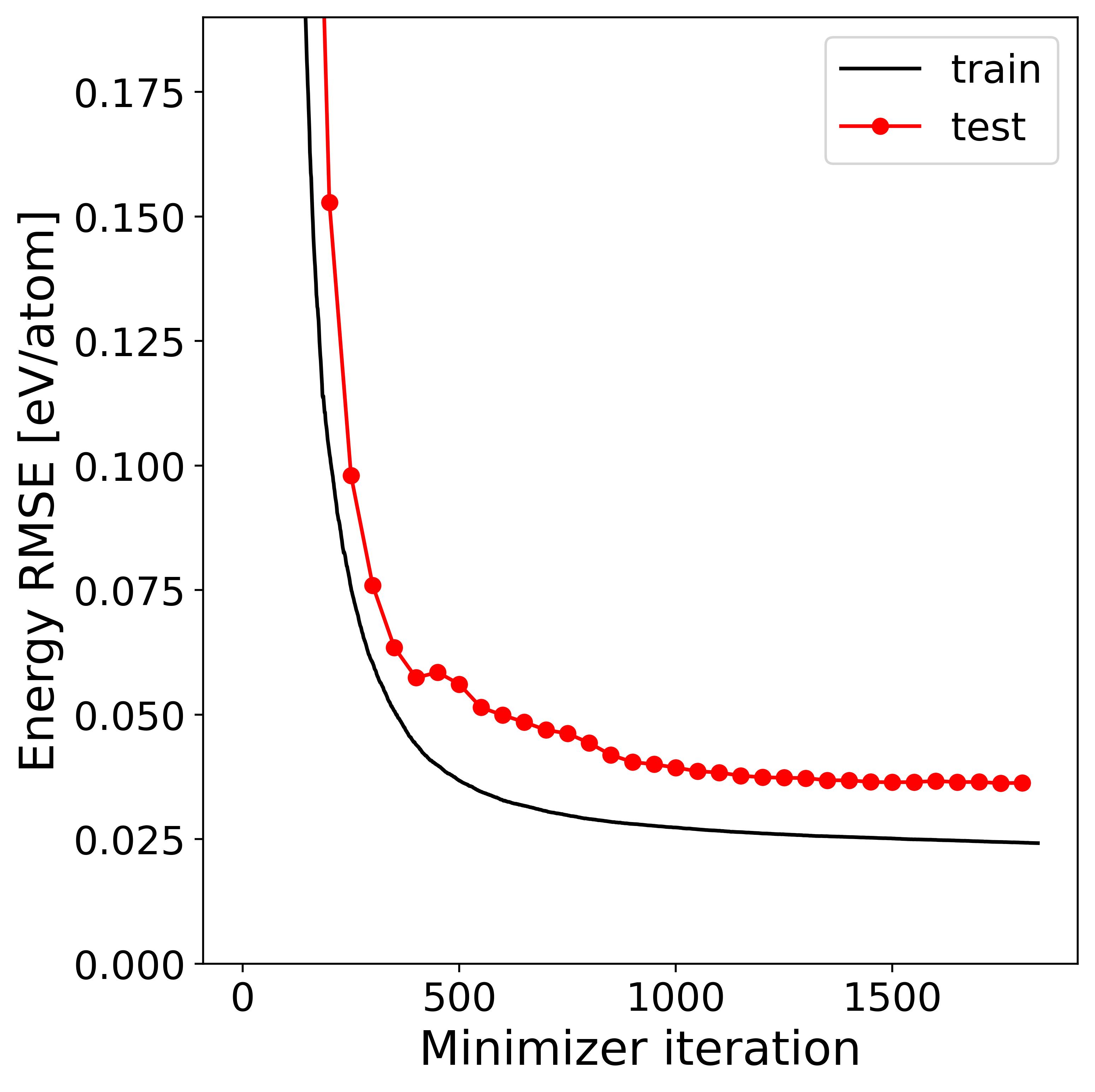}
    \caption{}
    \label{fig:RMSEEnergies}
  \end{subfigure}%
  \begin{subfigure}[t]{0.33\linewidth}
    \centering
    \includegraphics[width=0.99\linewidth]{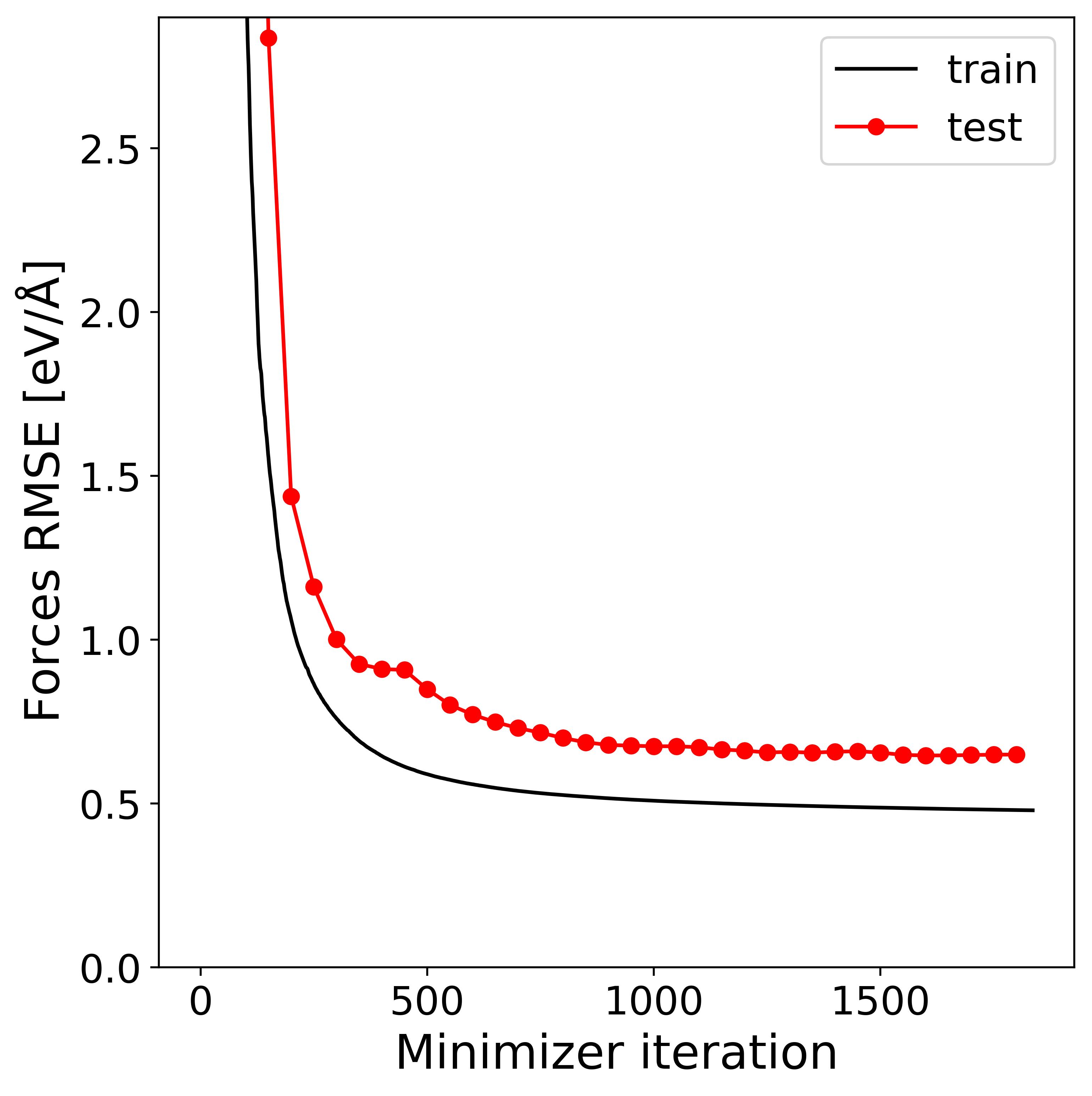}
    \caption{}
    \label{fig:RMSEForces}
  \end{subfigure}
  \caption{
    \textbf{Training data and performance of the potential.}
    Training data resulting from the \ac{AL} procedure (\subref{fig:EnergyVol}) with the black box indicating
    the magnified region shown in (\subref{fig:EnergyVolZoomed}).
    Error of energies as function of the formation energy distance from the convex hull (\subref{fig:ScatterEnergy})
    and correlation of forces (\subref{fig:ScatterForce}) as calculated with \ac{DFT} and the \ac{ACE} potential for the training and test datasets.
    Training and test RMSEs of energies (\subref{fig:RMSEEnergies}) and forces (\subref{fig:RMSEEnergies}) during optimization of the potential.
  }
  \label{fig:ScatterPlots}
\end{figure*}

\section{Results}

In the following we will start by shortly presenting the results of the applied \ac{AL}
procedure and evaluate the newly developed \ac{MLIP}.
Then we will discuss the structure of the produced \ac{SiOC} samples based on their composition and precursors
going from low to high C contents.
Finally, we will relate structural features to the Young's modulus of the samples.

\subsection{Training and performance of the potential}

The flexibility of the \ac{ACE} formalism allows for an accurate description
of highly complex materials,
under the condition that similar atomic configurations have
been part of the data used in their training procedure.
In this work, an \ac{AL} procedure
was employed to ensure that the training data covers a large phase space volume.
Here, we started with \acp{MTP} as implemented in the \MLIP{} package \cite{novikovMLIPPackageMoment2021},
and continued with \ac{ACE} potentials
for which \ac{AL} capabilities were implemented only recently \cite{lysogorskiyActiveLearningStrategies2023}.
We included structures with varying compositions and at high temperatures and pressures,
as well as different defective structures that can form during \ac{MD} simulations.
This makes the potential applicable to a wide area of problems.
Fig. \ref{fig:EnergyVol} and  \ref{fig:EnergyVolZoomed} show the resulting distribution of the training data in terms of
atomic energies and volumes.

The mostly actively learned \ac{SiOC} structures are scattered widely in
this 2D representation of phase space.
The pure elements and SiC structures were mostly made by hand.
In combination with less compositional degrees of freedom this leads to a comparatively narrow distribution for them.
To prevent the occurrence of unphysically large forces on single atoms during \ac{MD} simulations
at high temperatures or pressures we found it helpful to add SiC structures with a high density,
leading to a second area with aggregated SiC structures around \SI{4}{eV/atom} above the convex hull.

Energies and forces predicted using the potential agree well with those calculated using DFT
over the whole range of structures,
as shown in the scatter plots in Fig. \ref{fig:ScatterEnergy} and \ref{fig:ScatterForce}.
For energies an RMSE of \num{24} and \SI{36}{meV/atom} for the training and test sets were obtained.
For forces the RMSEs were \num{479} and \SI{650}{meV/\angstrom} respectively.
We want to note that the higher errors for the testing set
are a result from the different distribution of the structures regarding their composition
and energy and not from overfitting as shown by the continuous decrease of testing errors in Fig. \ref{fig:RMSEEnergies} and \ref{fig:RMSEForces}.

\subsection{Structure and energetics}

\subsubsection{\stoicho{}}

\begin{figure*}[tb!]
  \centering
  \includegraphics[width=1.0\linewidth]{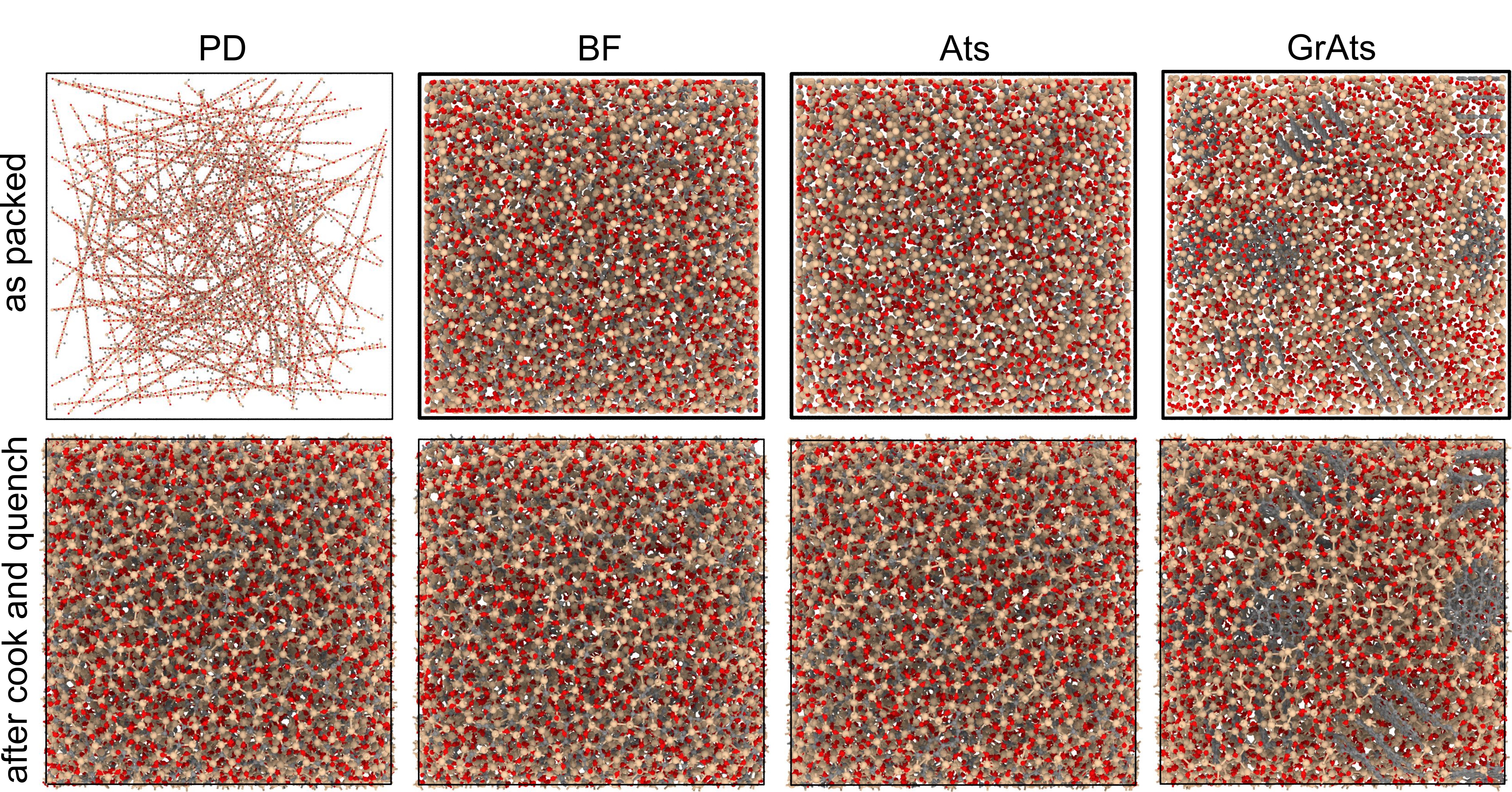}
  \caption{\textbf{\stoicho{} assembled precursors and final structures.}
    Shown are samples as packed with \PACKMOL{} (upper row) and after the cook and quench process (lower row) with an annealing temperature of \SI{1500}{K} for all structure prototypes.
    The PD, BF and Ats structures appear to be very similar upon visual inspection.
    The graphite like flakes in the GrAts based structure are still present after processing,
    showing that they are kinetically stabilized, as one would expect a decomposition into SiO$_2$ and SiC from an energetic viewpoint.
  }
  \label{fig:StructuresPackedProcessed}
\end{figure*}

The microstructure and properties of \ac{SiOC} compounds
depend strongly on the precursor material and processing conditions like pyrolysis temperature \cite{stablerSiliconOxycarbideGlasses2018}.
Time and lengths scales of the experimental processes can not be reproduced directly in \ac{MD} simulations.
Instead, we used different building blocks to produce a variety of microstructures
in cook and quench simulations as previously described.
Examples of these initial structures are shown in the upper row of Fig. \ref{fig:StructuresPackedProcessed}
Exemplary samples produced via the cook and quench protocol described in section \ref{sec:SampleStructures}
with an annealing temperature of \SI{1500}{K}
are shown in the row below.
Qualitatively, the \ac{PD}, \ac{BF} and \ac{Ats} structures
are very similar.
Major differences are only found for \ac{GrAts}.
Here, large graphite areas are still present in the final structure.
From a purely thermodynamic viewpoint this is surprising,
as one would expect that \stoicho{}
decomposes into SiO$_2$ and SiC,
but at the tested temperatures and timescales
the kinetics do not allow for such a phase separation.

\begin{figure*}[p]
  \centering
  \includegraphics[width=0.95\linewidth]{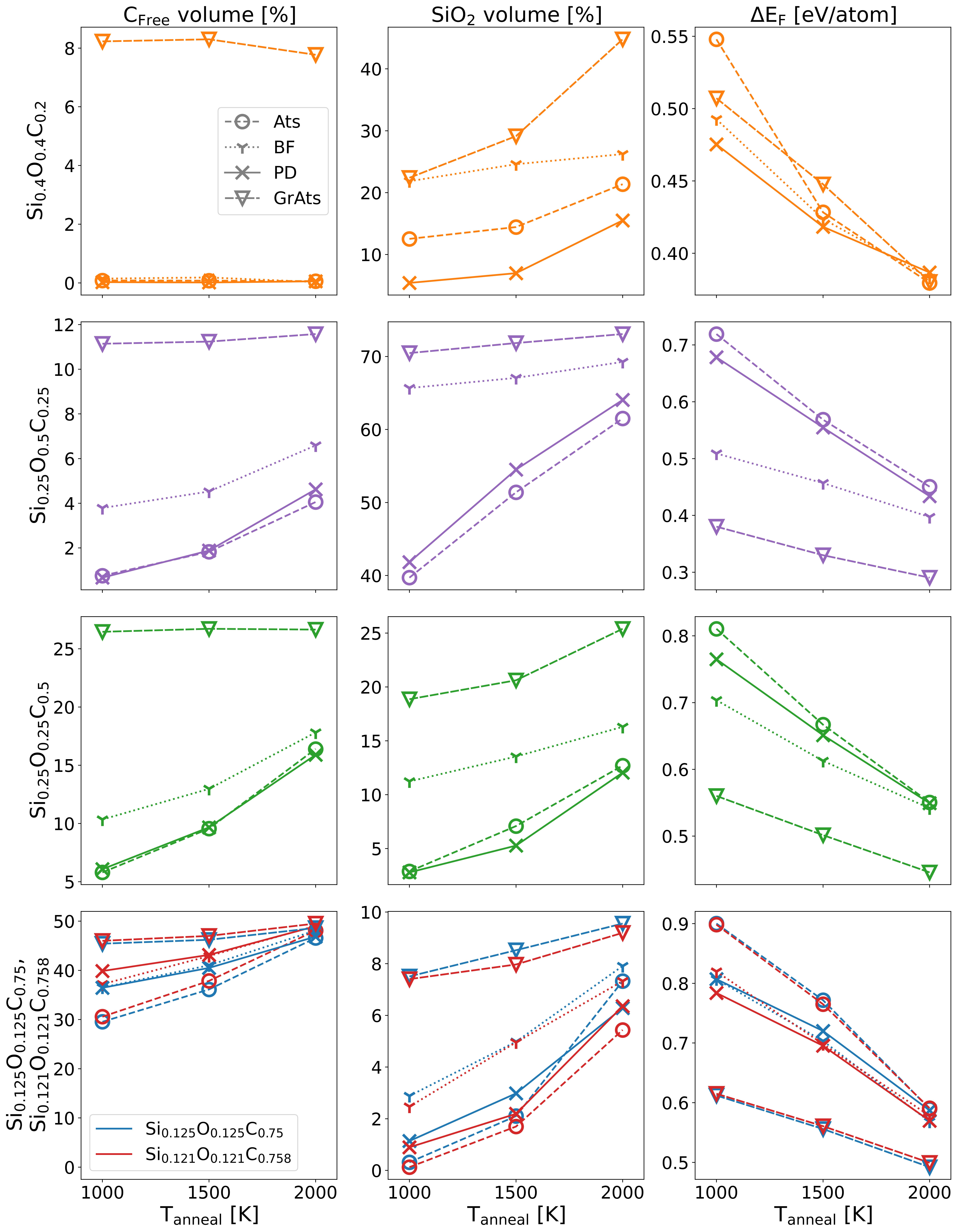}
  \caption{\textbf{Structure analysis and formation energies of \ac{SiOC} samples.}
    The left column shows the free carbon volume fraction, defined as Voronoi volume of all 3-fold coordinated C atoms, the center column shows the silica volume fraction (Voronoi volume of SiO$_4$ tetrahedra.
    The right column contains the formation energies of samples with respect to $\alpha$-quartz, $\beta$-SiC and graphite. Rows correspond to different sample compositions.
    \silres{} and \rdsix{} are shown in the same row, as a direct comparison of them is interesting especially in the case of PD structures due to their compositional similarity.}
  \label{fig:StructureAnalysis}
\end{figure*}

To quantify differences between
the structures we calculated the Voronoi volume fractions occupied by
3-fold coordinated C atoms, i.e. the free carbon phase
and SiO$_4$ tetrahedra, i.e. the silica phase.
The results are shown in Fig. \ref{fig:StructureAnalysis}.
Only the structures based on \ac{GrAts} contain a meaningful amount of free carbon.
As discussed previously there is no thermodynamic driving force for the formation of graphene
or graphite like carbon,
so its volume fraction is determined by the kinetics of the system and this behavior can be expected.
In the \ac{GrAts} structure the volume fraction stays constant between annealing temperatures \num{1000} and \SI{1500}{K}.
At \SI{2000}{K} a slight decrease can be observed, showing that the graphite phase is kinetically stabilized up to Temperatures greater than \SI{1500}{K}.
Regarding the silica volume fraction a continuous increase at increasing temperatures can be observed for all structures,
indicating a higher degree of phase separation.
However, the data shows large differences with regard to the absolute values.
The largest amount is observed for the \ac{GrAts} structure,
followed by \acp{BF}, \ac{Ats} and finally
the \ac{PD} configuration.
Again this observation can be rationalized by the thermodynamics and kinetics of the system.
In the \ac{GrAts} structure the unpaired Si and O atoms can quickly form
silica without interference of the carbon, because it is already bound in the graphite phase.
Si-C can only form in the relatively small interface regions or requires long range diffusion.
In the \ac{BF} based structure SiO$_2$, SiC$_4$ and Si$_4$C units are already present,
only requiring rotations and minor rearrangement to form the thermodynamically favored products,
while the diffusion paths necessary to achieve phase separation in the atom based structure
are long and intermediate bonds can form and need to brake again.
Similarly, the \ac{PD} structure contains Si bonded to C and O,
which needs to break before forming pure SiO$_4$ or SiC$_4$ tetrahedra.

The right column of Fig. \ref{fig:StructureAnalysis} shows the formation energy of samples with respect to $\alpha$-quartz, $\beta$-SiC and graphite.
In the case of \stoicho{} they are very close in energy, despite the significant structural differences.
Especially for the structures annealed at \SI{2000}{K} the difference of nearly 0 for may irritate,
when considering that the system still contains a considerable amount of free carbon in the case \ac{GrAts},
but this energetically unfavorable state is apparently compensated by the high fraction of very favorable silica in the system.

\subsubsection{\pmsq{}}

The carbon content of \pmsq{} is very close to the previously discussed \stoicho{},
but the different Si:O ratio leads to a different thermodynamic situation.
In equilibrium \pmsq{} should split into a pure silica and a graphite phase,
while no SiC should form.
Consequently, the formation energies of structures based on \ac{GrAts}
are considerably lower than those of the other precursors for \pmsq{}.
The difference between the \ac{Ats} and \ac{GrAts} structures
with an annealing temperature of \SI{1000}{K} is about \SI{300}{meV/atom},
i.e. the driving forces for phase separation are very high.
Since the structures are in a steady state, this also indicates a very low mobility.
For an annealing temperature of \SI{2000}{K} the difference between precursors becomes
much smaller, reaching around \SI{100}{meV/atom}.
It is expected that the differences shrink, because the increased mobility at higher temperatures
allows coming closer to the thermodynamic equilibrium.
The free carbon and silica volume in \pmsq{} samples increases significantly with increasing annealing temperatures.
The least changes are observed for the \ac{GrAts} structure,
which is already close to the phase separation.
Generally, the arguments regarding the kinetics of the different precursors discussed for \stoicho{} also apply
for \pmsq{}, so similar trends can be observed,
with the main differences determined by the decomposition products.

\subsubsection{\rdtwo{}, \silres{} and \rdsix{}}

\rdtwo{}, \silres{} and \rdsix{}
contain Si and O in a 1:1 ratio,
but, in contrast to \stoicho{}, also excess carbon.
Thus we expect the formation of a varying amount of
free carbon phase in the system
and not just SiO$_2$ and SiC.
The compositional similarity of \silres{} and \rdsix{}
allows us to investigate the influence of different polymer-like precursor structures.
In the \rdsix{} polymer C$_6$ rings are stacked very closely (cf. Fig. \ref{fig:BuildingBlock}),
which could ease the formation of graphite.
However, this can not be observed in our data.
Generally the three compounds show high free carbon contents
and qualitatively similar temperature dependence for both,
the silica and free carbon fractions.
The energetically most favorable sample is the one made from
\ac{GrAts}.
Similar to \stoicho{} this can be explained by the high degree of phase separation,
which corresponds to the thermodynamic equilibrium.
\silres{} and \rdsix{} behave similar.

\begin{figure*}[tbp!]
  \centering
  \includegraphics[width=\linewidth]{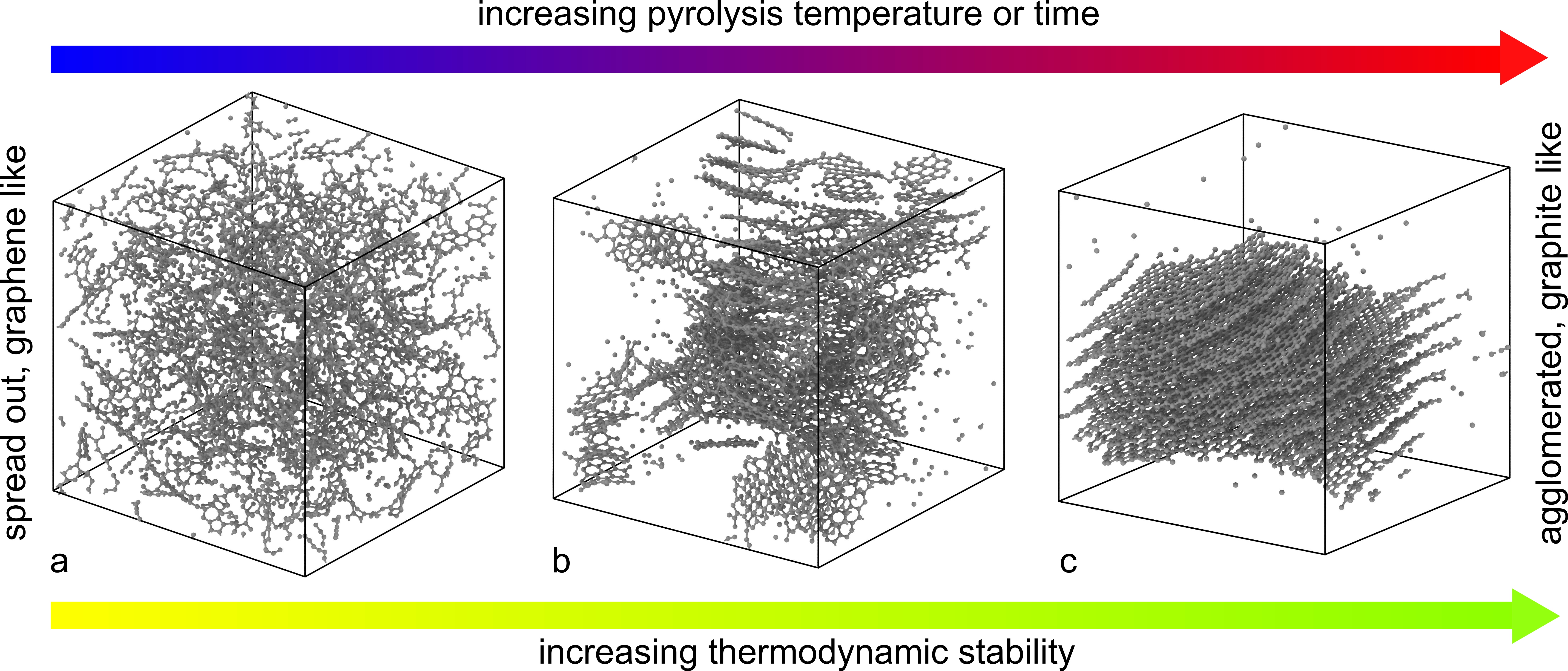}
  \caption{\textbf{Structure formation of free carbon phase in \ac{SiOC}.}
    For low pyrolysis temperatures and times graphene like carbon forms as argued by Scarmi et al. \cite{scarmiRoleCarbonUnexpected2005} and Saha et al. \cite{sahaModelNanodomainsPolymerDerived2006} based on high creep resistance.
    At higher temperatures energetically more favorable graphite like agglomerates form, which are consistent with the mass fractal dimension as found by Widgeon et al. \cite{widgeon29Si13CSolidState2010}.
    The figures show the free carbon phase as found for \rdtwo \ac{PD} sample annealed at \SI{2000}{K} (a), \ac{GrAts} sample annealed at \SI{2000}{K} (b) and an additional \ac{PD} sample annealed at \SI{3500}{K} (c).
    Their formation energies suggests that the \ac{SiOC} system would evolve from \ac{PD} to \ac{GrAts}
    structures, so we consider it as an intermediate step even though timescales accessible in \ac{MD} do not allow a direct observation.
  }
  \label{fig:SiOCModelStructures}
\end{figure*}

\subsubsection{Relation to model structures}

As discussed previously, two models for the nanostructure of
\ac{SiOC} exist.
One suggests silica rich nanodomains, which are separated by
an interconnected graphene like carbon network \cite{scarmiRoleCarbonUnexpected2005,sahaModelNanodomainsPolymerDerived2006}.
The other describes the structure as graphitic inclusions in a silica rich matrix  \cite{widgeon29Si13CSolidState2010}.
The results from our simulations are shown in Fig. \ref{fig:SiOCModelStructures}.
For structures that should contain graphite upon decomposition we found
the \ac{GrAts} models to be energetically favorable,
favoring the latter model.
However, in the other samples no graphitic carbon could be found
and instead graphene like layers spread through the system,
pointing to the former.
This allows us to conclude that both models are representative for two distinct stages.
In early stages of structure formation the structure is likely described by the first model,
because the slow kinetics in the system prevent the formation of
graphitic inclusions.
With increasing temperatures and pyrolysis times
the structure evolves towards
the latter model argued for by Widgeon et al.,
as it is lower in energy.

\begin{figure*}[tb!]
  \centering
  \begin{subfigure}{0.4\linewidth}
    \centering
    \includegraphics[width=\linewidth]{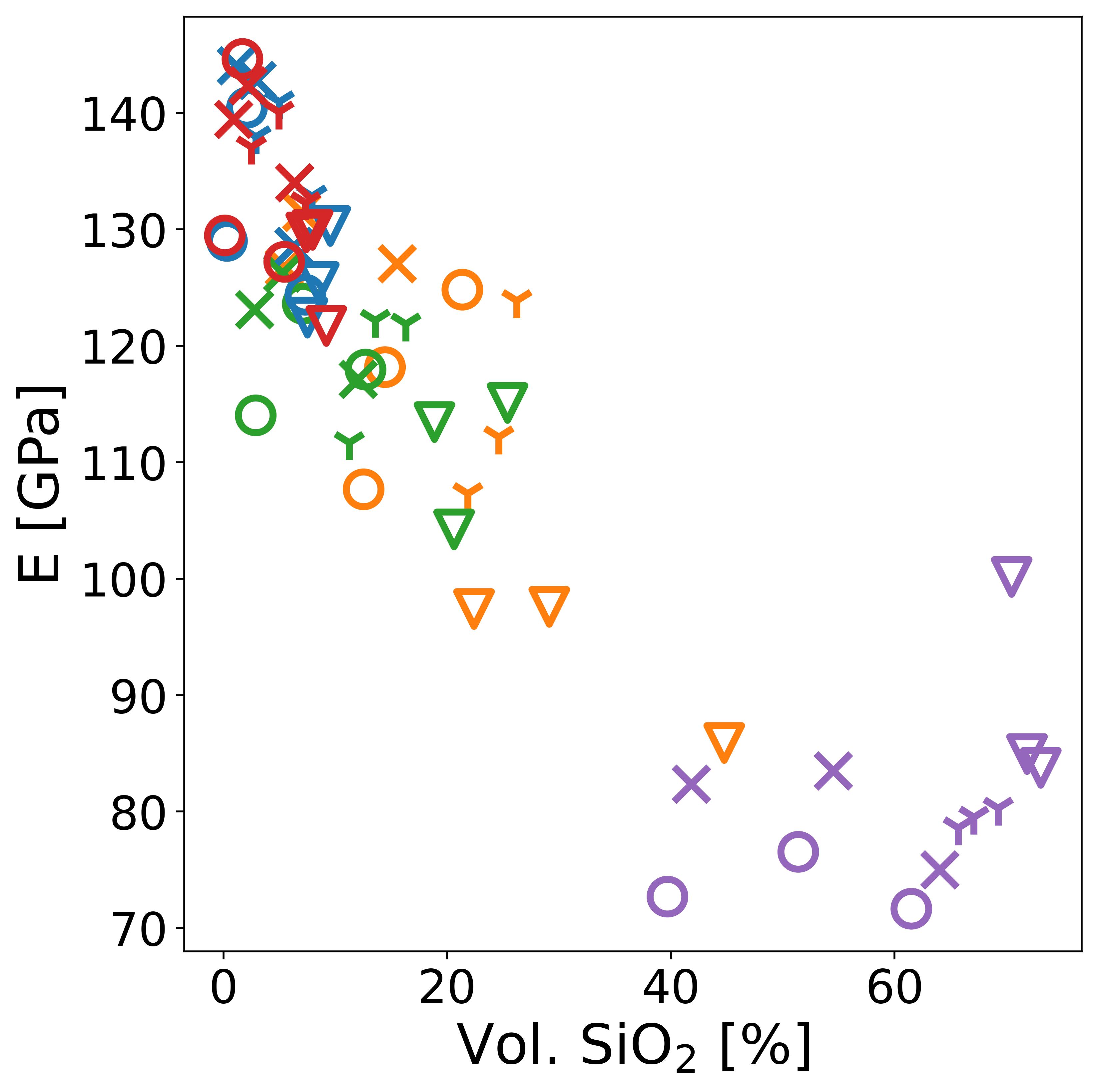}
    \caption{}
    \label{fig:ESilica}
  \end{subfigure}%
  \begin{subfigure}{0.4\linewidth}
    \centering
    \includegraphics[width=\linewidth]{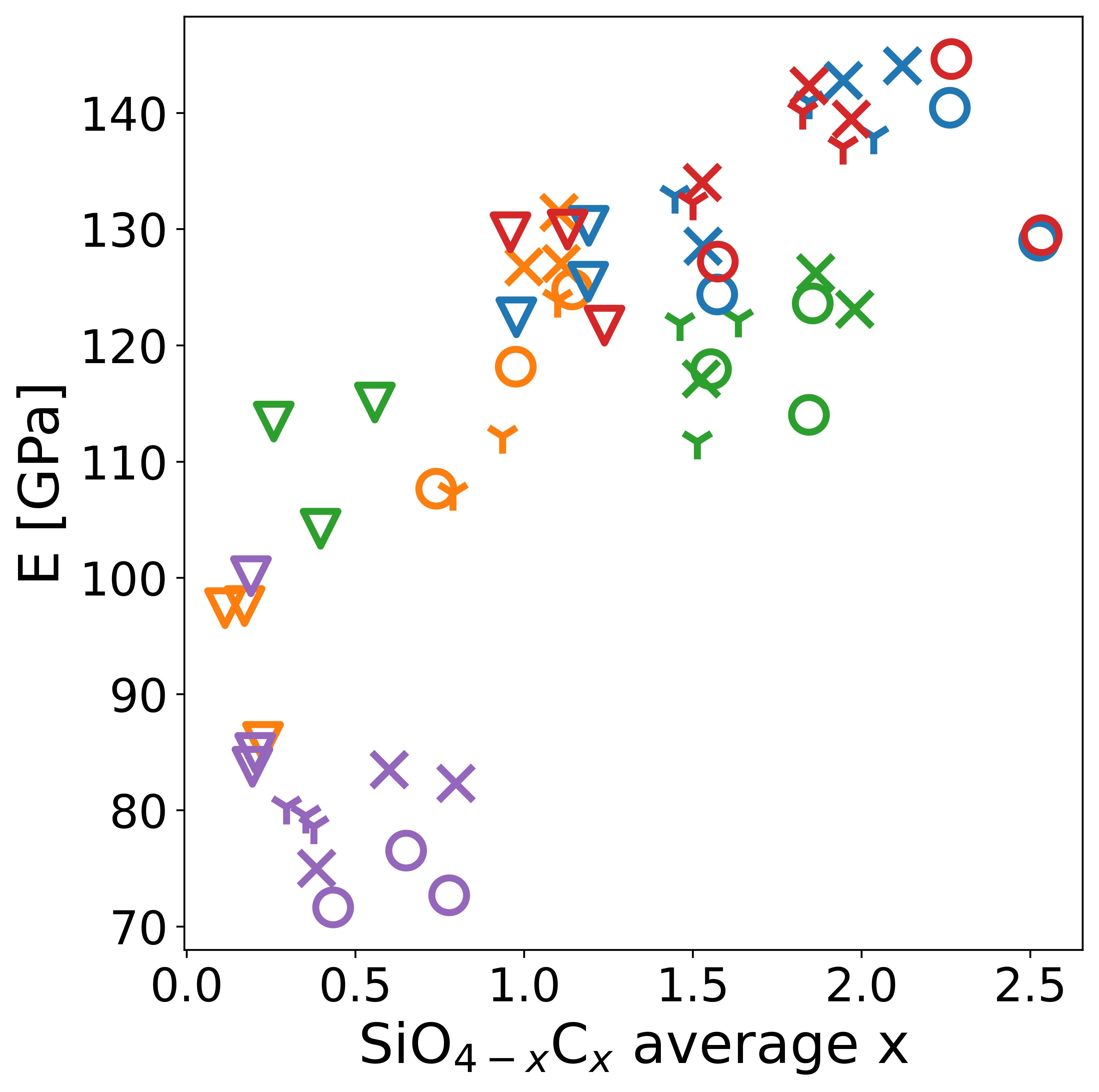}
    \caption{}
    \label{fig:EaverageX}
  \end{subfigure}
  \begin{subfigure}[t]{0.4\linewidth}
    \centering
    \includegraphics[width=\linewidth]{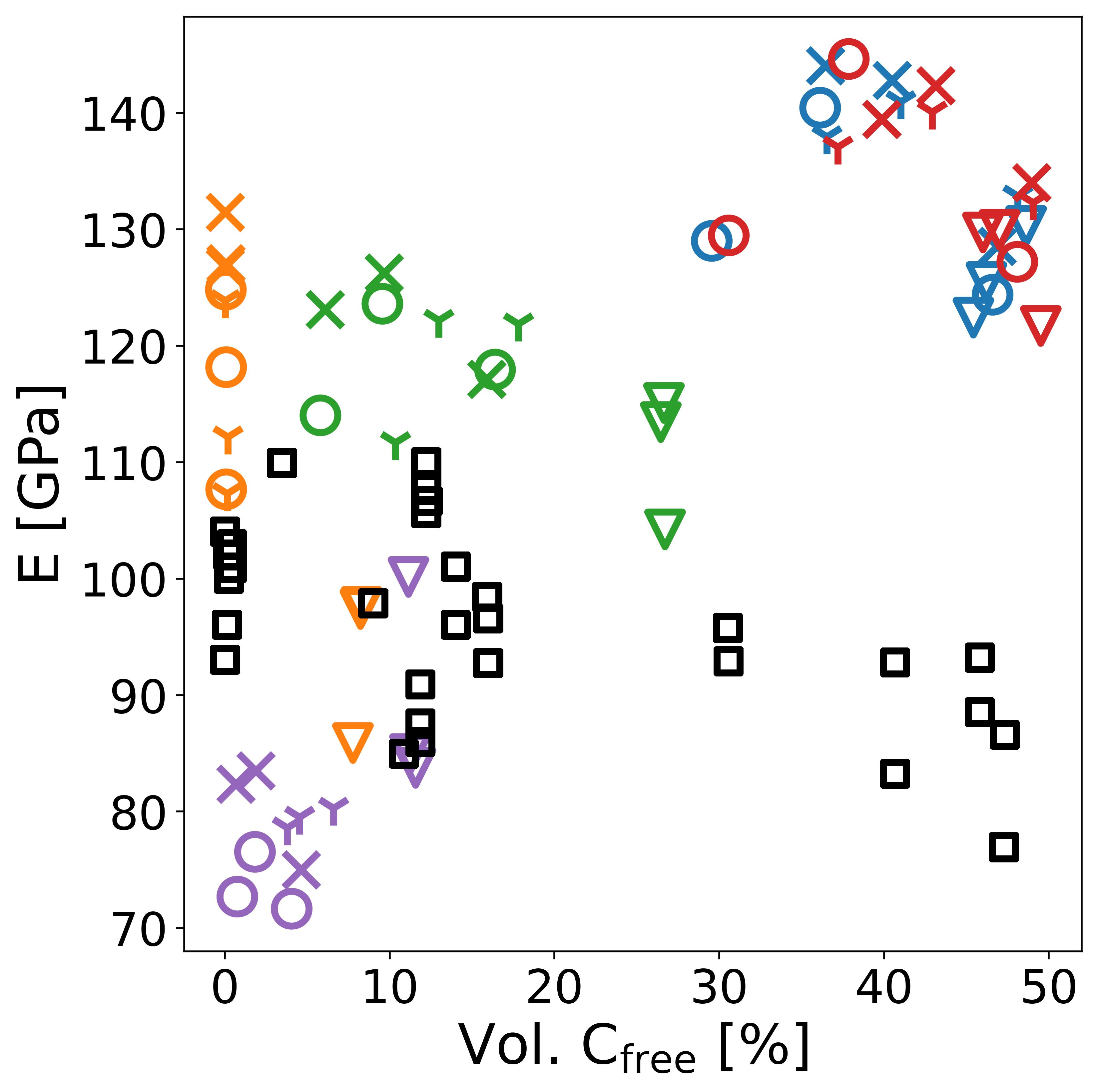}
    \caption{}
    \label{fig:EFreeCarbon}
  \end{subfigure}%
  \begin{subfigure}[t]{0.4\linewidth}
    \centering
    \includegraphics[width=0.5\linewidth]{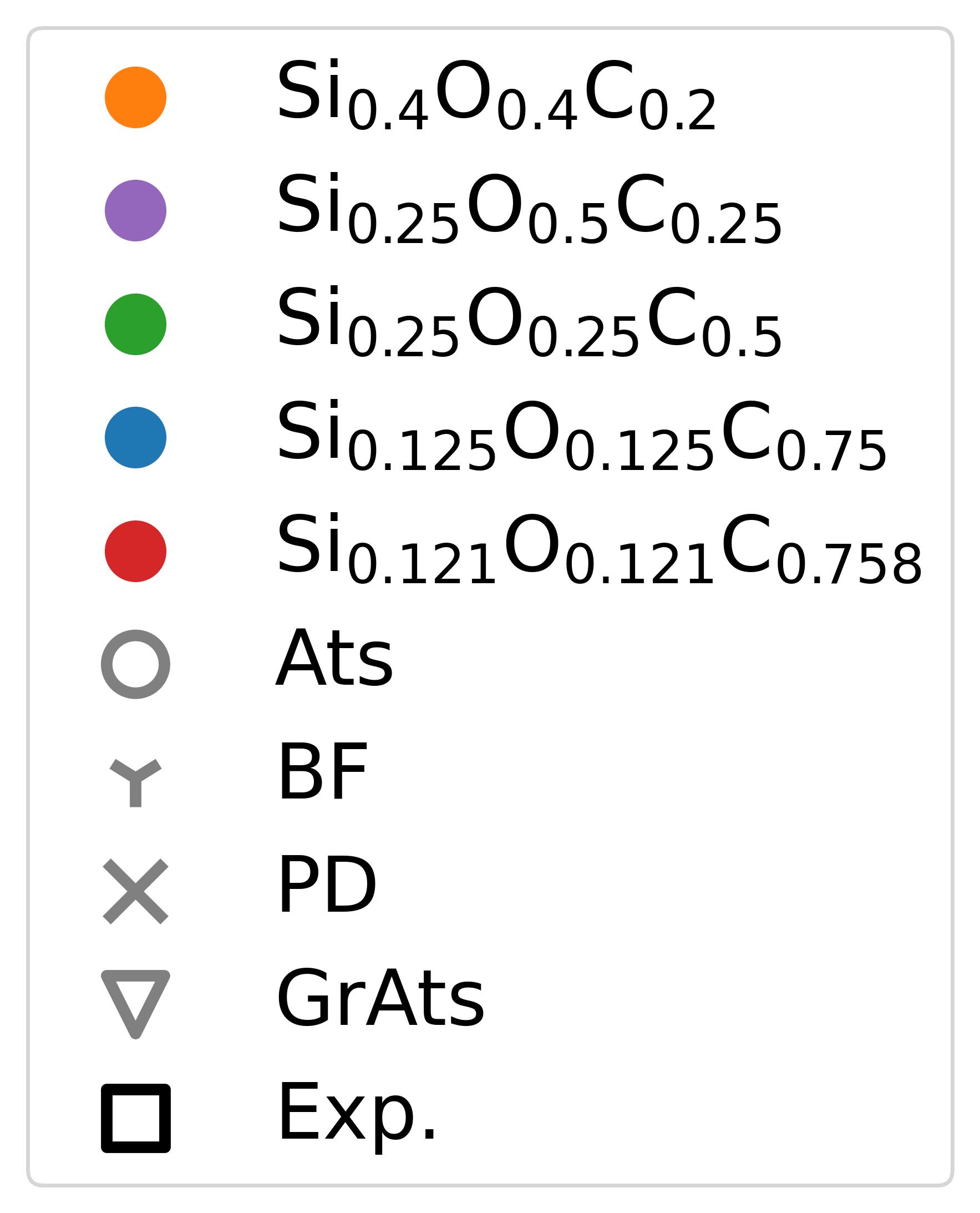}%
  \end{subfigure}%
  \caption{\textbf{Dependence of E on structural features}.
    Calculated E values for samples in dependence of silica volume fraction (\subref{fig:ESilica}), average amount of carbon in mixed tetrahedra (\subref{fig:EaverageX}) and free carbon volume fraction (\subref{fig:EFreeCarbon}).
    Experimental data in (\subref{fig:EFreeCarbon}) is taken from
    \cite{renlundSiliconOxycarbideGlasses1991,soraruMechanicalCharacterizationSol1996,walterMicrostructuralMechanicalCharacterization2002,moysanMechanicalCharacterizationPolysiloxanederived2007,papendorfHighTemperatureCreepBehavior2013,stablerEffectCompositionHightemperature2019}
    as collected in \cite{stablerEffectCompositionHightemperature2019} and from \cite{soraruInfluenceFreeCarbon2019}.}
  \label{fig:YoungsModulus}
\end{figure*}

\subsection{Elastic properties}

The structure of \ac{SiOC} is highly tunable, depending on the pyrolysis conditions and precursors.
An understanding of structure-property relations can therefore guide the search for processing routes that best match specific needs.
Here, we investigate the structural features influencing the elastic properties of \ac{SiOC}.
For this purpose we calculated the elastic tensors of our samples using pymatgen \cite{ongPythonMaterialsGenomics2013} and derived
the \acl{E}
\begin{equation}
  E = \mu (3\lambda + 2\mu) / (\lambda + \mu)
\end{equation}
with the Lamè constants
$\mu = C_{44}$ and $\lambda = C_{12}$,
as valid for isotropic materials.
Here, the assumption of isotropy is very accurate with differences between $C_{ij}$ that should be equivalent due to symmetry in the range of \SI{\pm 3}{\%}.
Only the \ac{GrAts} structures show larger anisotropies,
due to the limited amount of randomly orientated graphite flakes,
leading to differences in the range of \SI{\pm 15}{\%}.
To reduce resulting errors we averaged over the supposedly equivalent directions.
The deformations applied to obtain elastic tensors are not fully reversible because the relaxation of atomic positions
leads to small energy barriers on the path back to the initial state that cannot be overcome in static relaxations.
Corresponding energy-strain relations can be found in the supplemental material.

Fig. \ref{fig:YoungsModulus} shows the dependence of \ac{E} on the SiO$_4$ volume,
the average amount of carbon in SiO$_{4-x}$C$_x$ tetrahedra and the free carbon volume
in the samples.
Generally the modulus ranges from \SI{70}{GPa},
which corresponds to \ac{E} of silica glass \cite{inabaYoungModulusCompositional1999}
up to around \SI{145}{GPa}, which agrees well with experiments.
Also, it scatters significantly between different compositions and building blocks,
showing the multidimensional nature of the problem.
Strong correlations of \ac{E} can be seen for the SiO$_4$ volume fraction and the amount of carbon within SiO$_{4-x}$C$_x$ tetrahedra.
As expected the modulus decreases for higher silica volume fractions,
approaching that of the pure amorphous phase and
increases with an increasing amount of carbon in the mixed bonds.
As discussed previously, \ac{GrAts} based structures contain high amounts of SiO$_4$ tetrahedra
and the C atoms are bound within the graphite inclusions,
i.e. they do not participate much in mixed bonding.
Consequently, these structures have lower Young's moduli than structures based on other building blocks
at the same compositions.
Experimental studies have established
empirical relations between structural
features of \ac{SiOC} and its elastic properties.
Typically, it is observed that the Young's modulus
increases with the amount of SiC and C within the SiO$_{4-x}$C$_x$ tetrahedra and decreases
with increasing free carbon volume
\cite{soraruMechanicalCharacterizationSol1996,soraruInfluenceFreeCarbon2019}.
However, as shown by Stabler et al. \cite{stablerEffectCompositionHightemperature2019}
the results can differ significantly based on the kind of sample and applied measurement method.
Furthermore, a recent simulation based study by
Haseen and Kroll \cite{haseenAnalyzingEffectComposition2023}
found a strong decrease of E with decreasing sample densities
employing a Tersoff type potential.
They argued that this could overlap the direct effect of the C$_\mathrm{free}$ phase,
because the density of samples also strongly correlates with the amount of C$_\mathrm{free}$.
Indeed, they observed an increase of E with higher free carbon contents for samples with same density.
These results fit well to our observation of a weak overall correlation of E with C$_\mathrm{free}$.
If anything, when taking into account all data points we observe a slight increase.
Considering each composition individually can also help to resolve the discrepancy with experimental results, as all of them but \pmsq{} show downwards trends.
Therefore, we conclude that the amount of free carbon only weakly influences \ac{E},
but at similar compositions more free carbon is equivalent to lower amounts of the strong Si-C bonds,
indirectly leading to a lower stiffness.

\section{Conclusion}

Using multiple iterations of \ac{AL} techniques implemented for \acp{MTP} and \ac{ACE} potentials we produced a highly diverse set
of amorphous \ac{SiOC} structures, spanning a large area of phase space.
With the converged dataset, we fitted a nonlinear \ac{ACE} potential to energies and forces
calculated with \ac{DFT} and showed that the potential can accurately reproduce them.
Due to the diversity of the training data,
the potential has a large applicability range and
can describe the formation of Si-O-C compounds and their properties
in a large temperature and pressure range.

Applying the potential, we produced amorphous \ac{SiOC} samples with various compositions
in a cook and quench procedure.
Here, we tested the influence of different initial structures on the final configuration.
We found that manually added graphite agglomerates are kinetically stabilized in \ac{MD} simulation times
and lead to energetically favorable states
compared to graphene like sheets, if the structure contains excess C.
However, their formation is kinetically hindered,
and we assume that an interpenetrating network of silica rich domains and graphene
like sheets is formed as an intermediate step during synthesis
of amorphous \ac{SiOC}.

To test and establish structure-property relations in \ac{SiOC} we
calculated Young's moduli of the samples.
We found that the silica volume fraction and the average amount of carbon in
mixed SiO$_{4-x}$C$_x$ tetrahedra correlate well with the stiffness across all samples.
Silica reduces the stiffness and high amounts of Si-C bonds increase it.
Furthermore, the free carbon volume fraction is a good indicator of \ac{E} for samples with similar compositions,
but not in general.

We believe that the presented potential will allow detailed investigations
of structure formation and properties in \ac{SiOC}.
In a broader context, we have shown that modern \acp{MLIP}
can be employed to study multi element material systems
that form highly complex structures.
Therefore, they can greatly benefit the understanding of
glassy and ceramic materials on the atomistic scale.

\section*{Acknowledgments}

N. L. acknowledges Linus C. Erhard for helpful discussions.
N. L. recieved funding from the Deutsche Forschungsgemeinschaft (DFG, German Research Foundation) under grant number 405621137 and from the German Federal Ministry of Education and Research (BMBF) under project HeNa (FKZ 03XP0390A). J. R. acknowledges funding from the European Unions Horizon 2020 research and innovation programme under Grant Agreement no. 963542.
K. A. acknowledges funding by the DFG under project number 413956820.
The authors gratefully acknowledge the computing time provided to them on the high-performance computer Lichtenberg at the NHR Centers NHR4CES at TU Darmstadt. This is funded by the German Federal Ministry of Education and Research, and the state governments participating.

\section*{Data availability}

The potential and datasets are available on zenodo upon journal acceptance \cite{leimerothPotentialTrainingData}.

\section*{Author contributions}

N. L. performed most computations and analysis.
J. R. provided hand crafted \ac{SiOC} structures and performed early \ac{AL} iterations with \acp{MTP}.
All authors contributed to the research design and interpretation of results.
N. L. wrote the paper with input from J. R. and K. A.

\section*{Competing interests}

The authors declare no competing interests.

\printbibliography

\end{document}


\maketitle

\setcounter{table}{0}
\renewcommand{\thetable}{S\arabic{table}}%
\setcounter{figure}{0}
\renewcommand{\thefigure}{S\arabic{figure}}%
\setcounter{section}{0}
\renewcommand{\thesection}{S\arabic{section}}%

\acrodef{DFT}[DFT]{density-functional theory}
\acrodef{MD}[MD]{molecular dynamics}
\acrodef{ACE}[ACE]{Atomic Cluster Expansion}
\acrodef{MTP}[MTP]{Moment Tensor potential}
\acrodef{MLIP}[MLIP]{machine learning interatomic potential}
\acrodef{IP}[IP]{Interatomic potential}
\acrodef{SOAP}[SOAP]{Smooth Overlap of Atomic Positions}
\acrodef{NNP}{Neural Network potential}
\acrodef{SiOC}[Si-O-C]{silicon oxycarbide}
\acrodef{AL}[AL]{active learning}
\acrodef{PMSQ}[PMSQ]{Polymethylsilsesquioxane} 

\section{Steady state of glassy structure\label{sec:SteadyState}}

\begin{figure*}[htbp!]
    \centering
    \begin{subfigure}{0.4\linewidth}
        \centering
        \includegraphics[width=\linewidth]{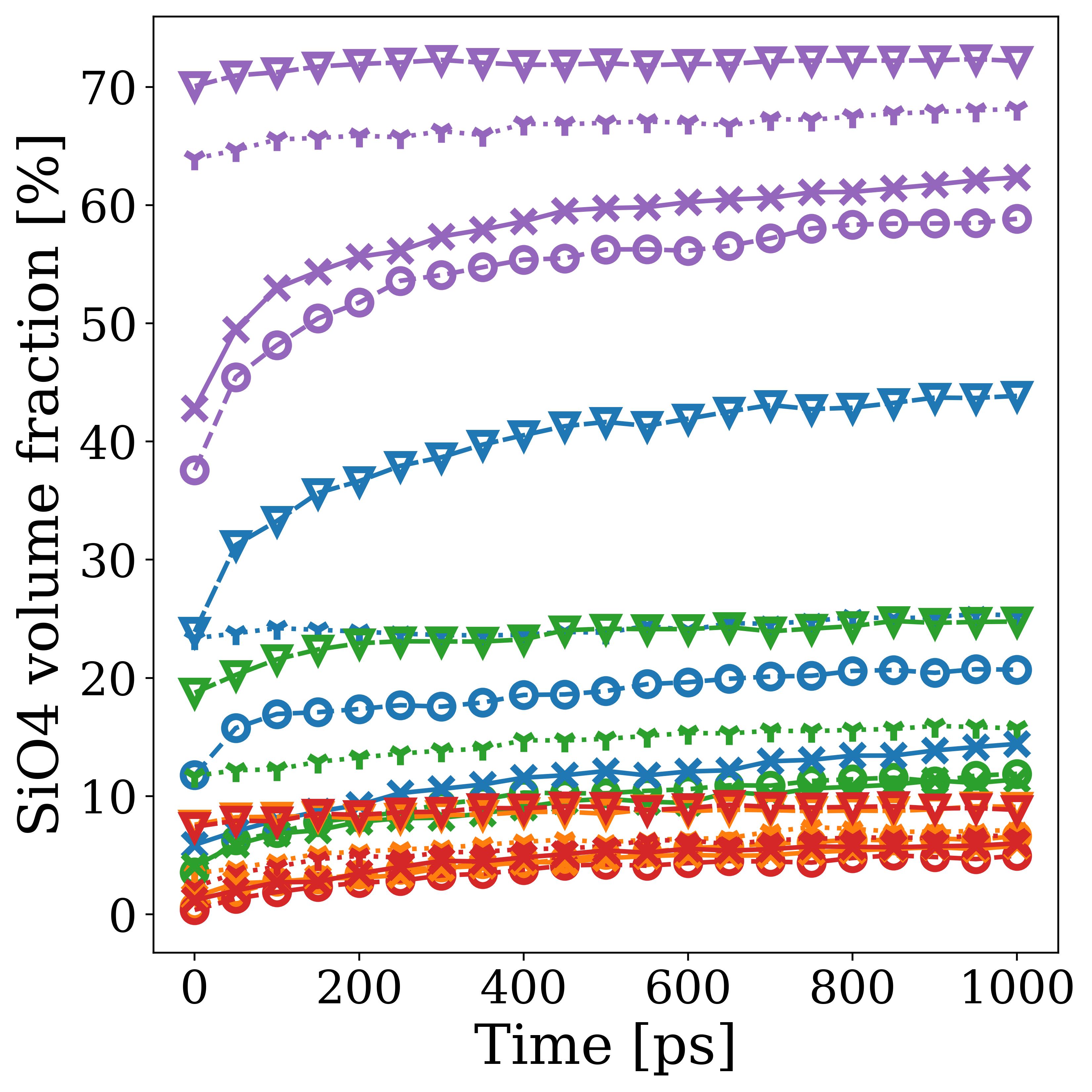}
    \end{subfigure}%
    \begin{subfigure}{0.4\linewidth}
        \centering
        \includegraphics[width=\linewidth]{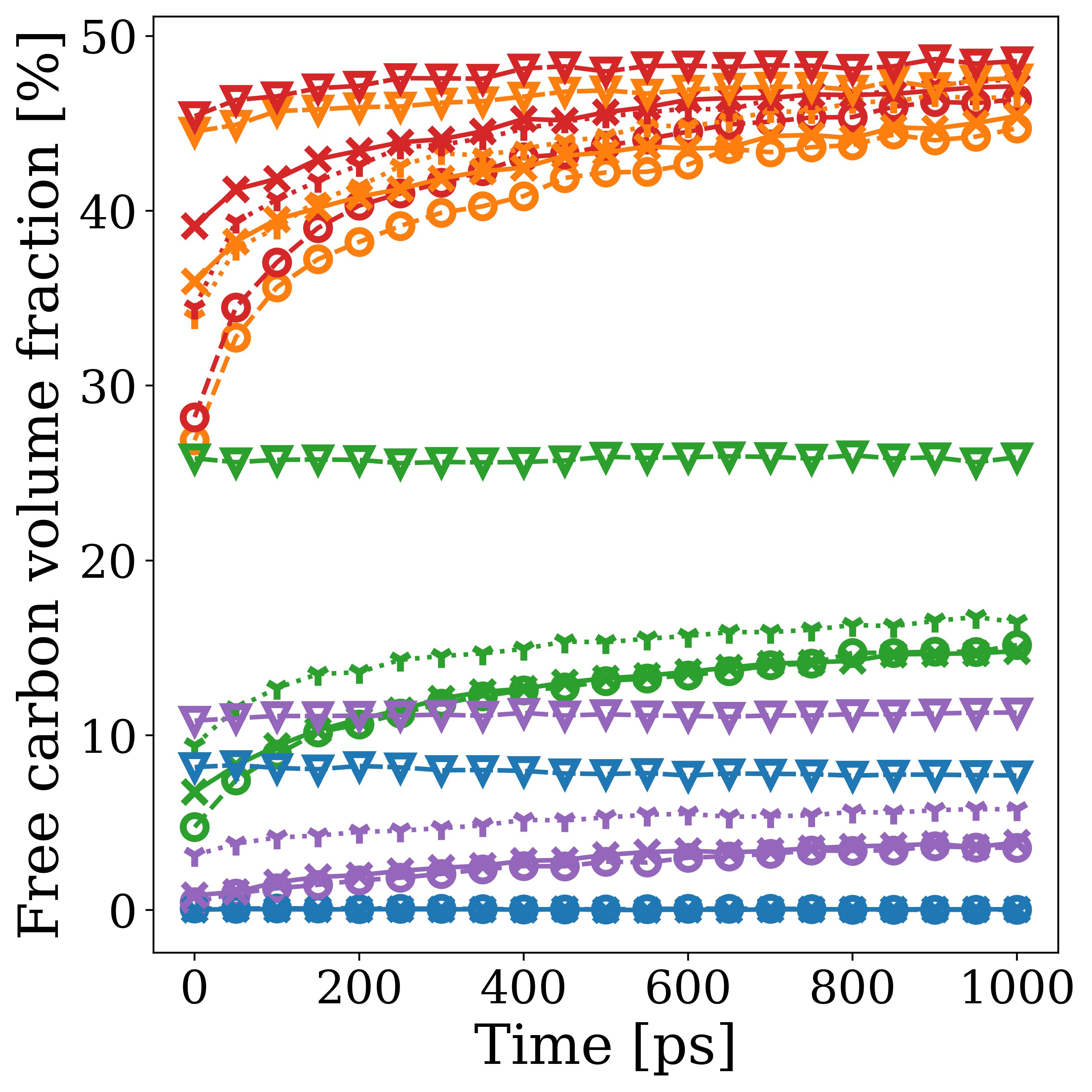}
    \end{subfigure}
    \begin{subfigure}{0.19\linewidth}
        \centering
        \includegraphics[width=\linewidth]{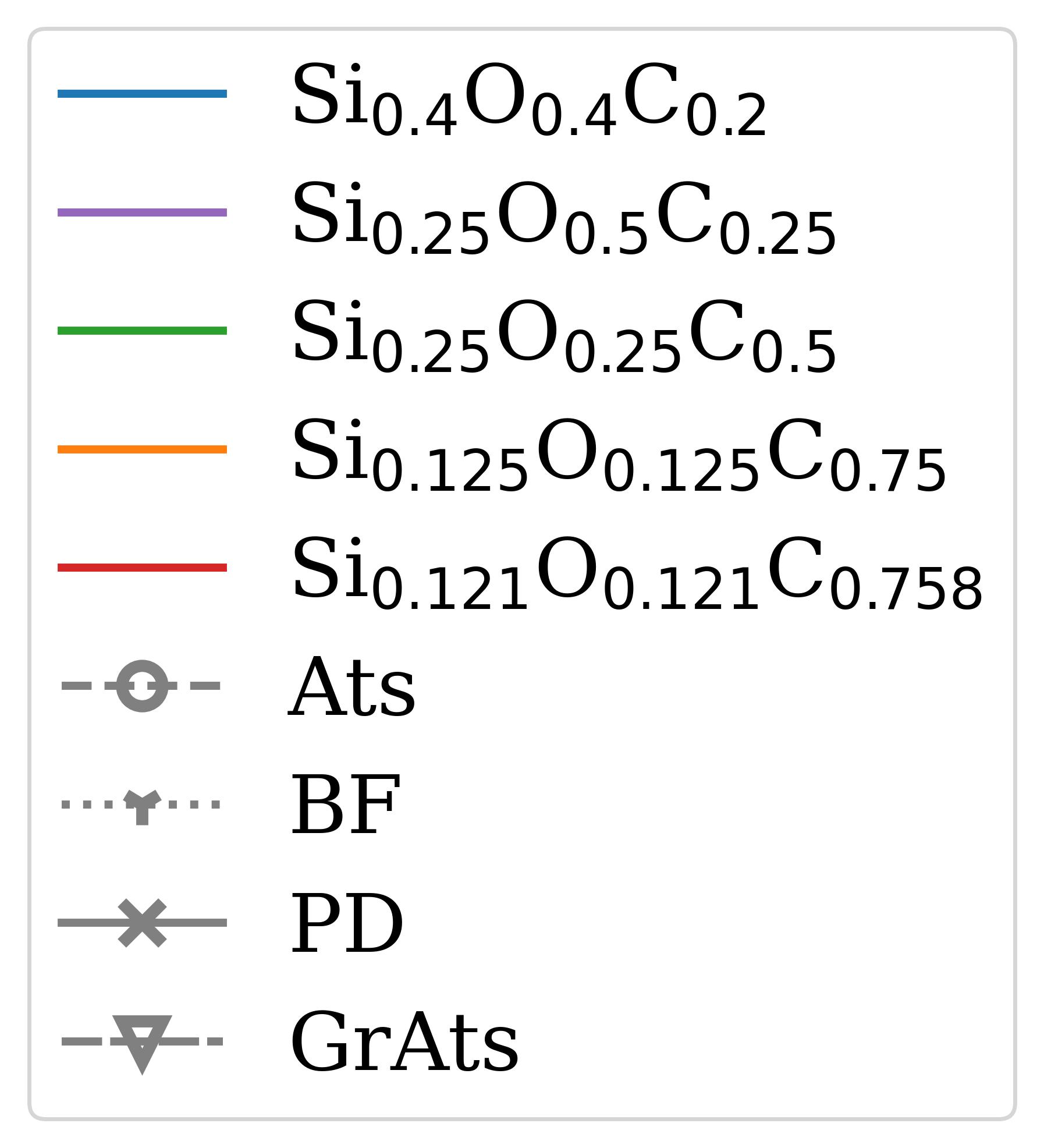}
    \end{subfigure}%
    \caption{Volume fraction of SiO$_4$ tetrahedra and free carbon while equilibrating at \SI{2000}{K}.}
    \label{fig:VolOverTime}
\end{figure*}

Fig. \ref{fig:VolOverTime} shows volume fractions of structural features
while equilibrating the structures during cook and quench simulations.
After \SI{1}{ns} the structural features do not undergo major changes anymore.

\section{Si centered tetrahedra \label{sec:SiCenteredTetrahedra}}
\begin{figure*}[htbp!]
    \centering
    \includegraphics[width=0.9\linewidth]{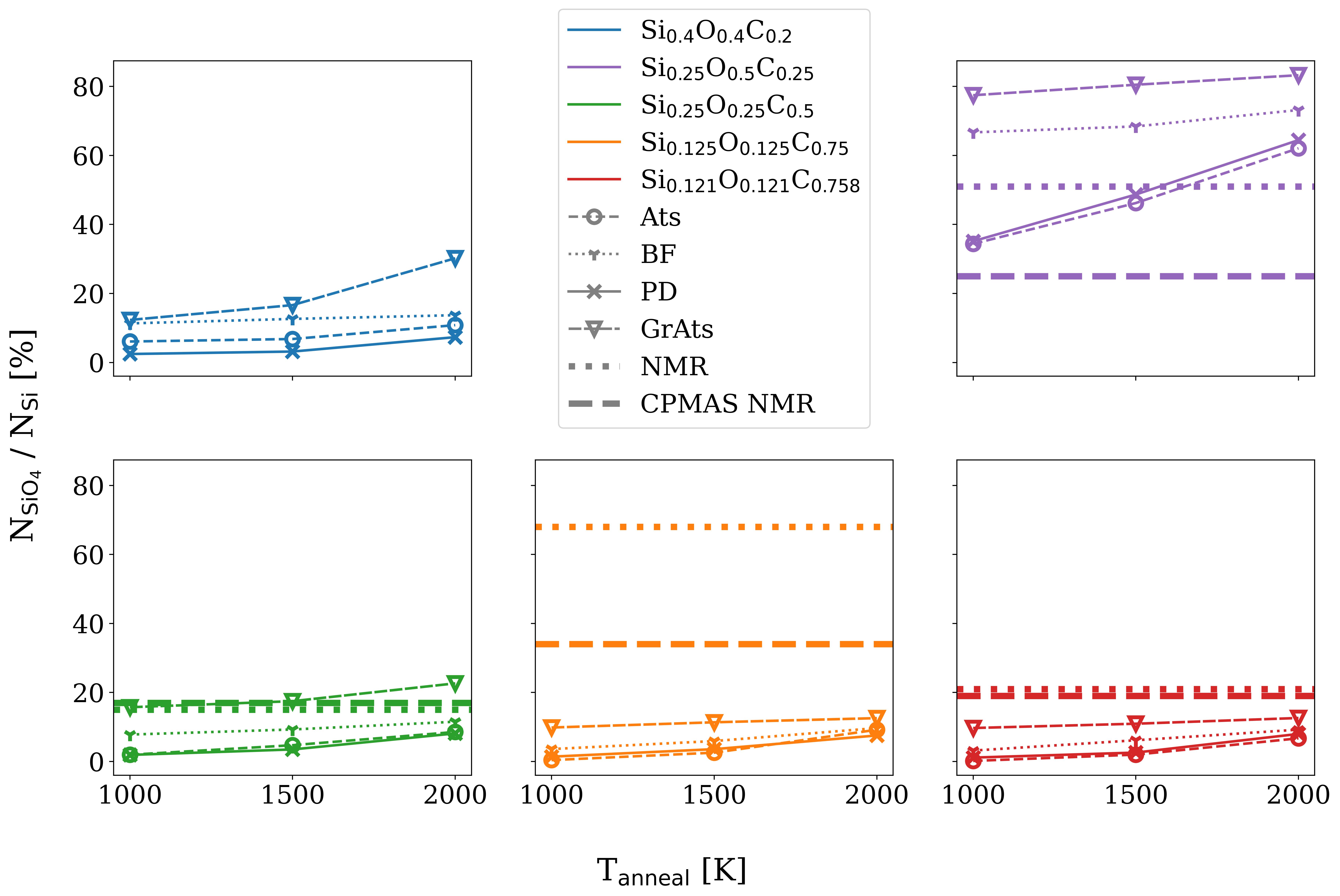}
    \caption{See text for more information.}
    \label{fig:SiO4Frac}
\end{figure*}
\begin{figure*}[htbp!]
    \centering
    \includegraphics[width=0.9\linewidth]{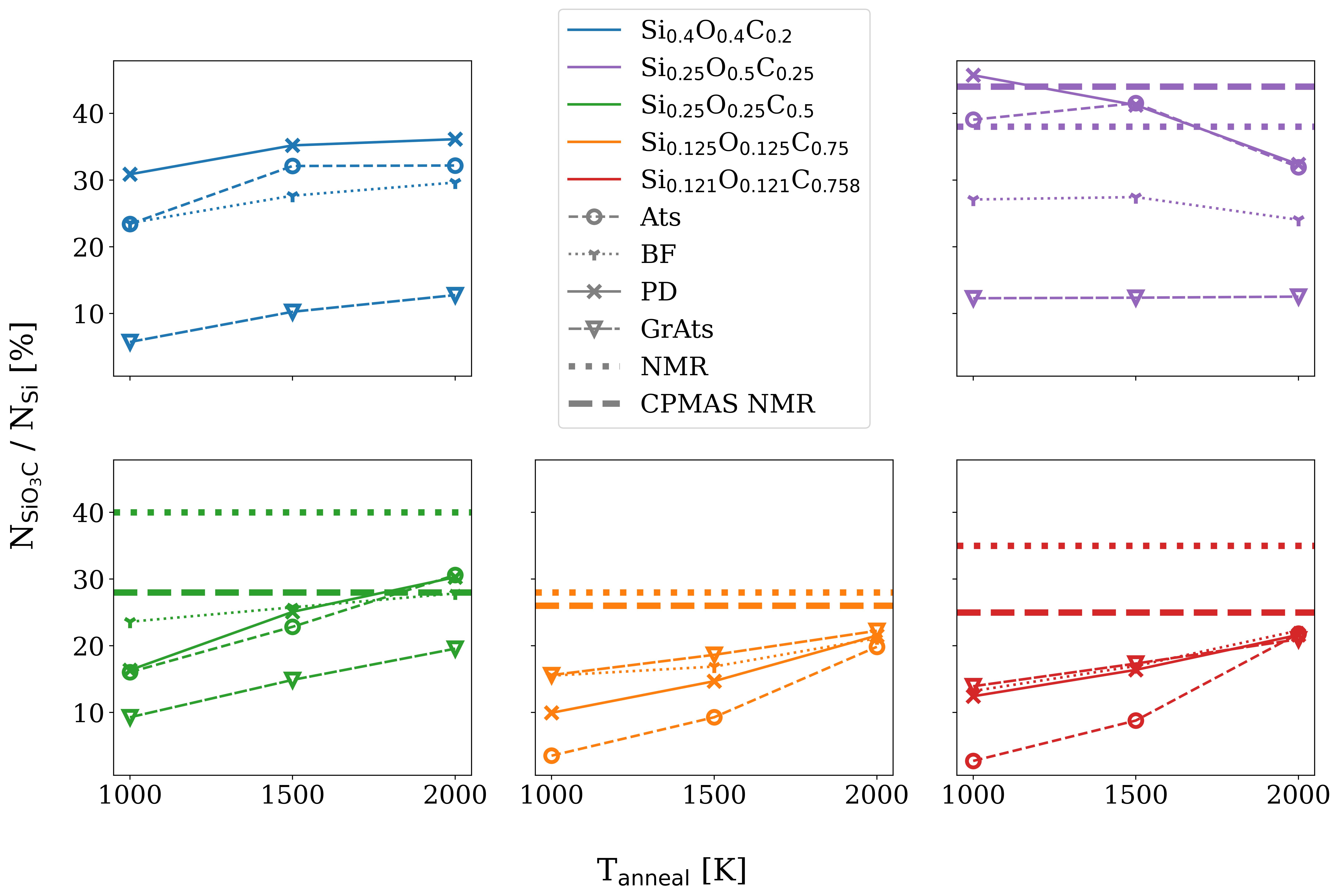}
    \caption{See text for more information}
    \label{fig:SiO3CFrac}
\end{figure*}
\begin{figure*}[htbp!]
    \centering
    \includegraphics[width=0.9\linewidth]{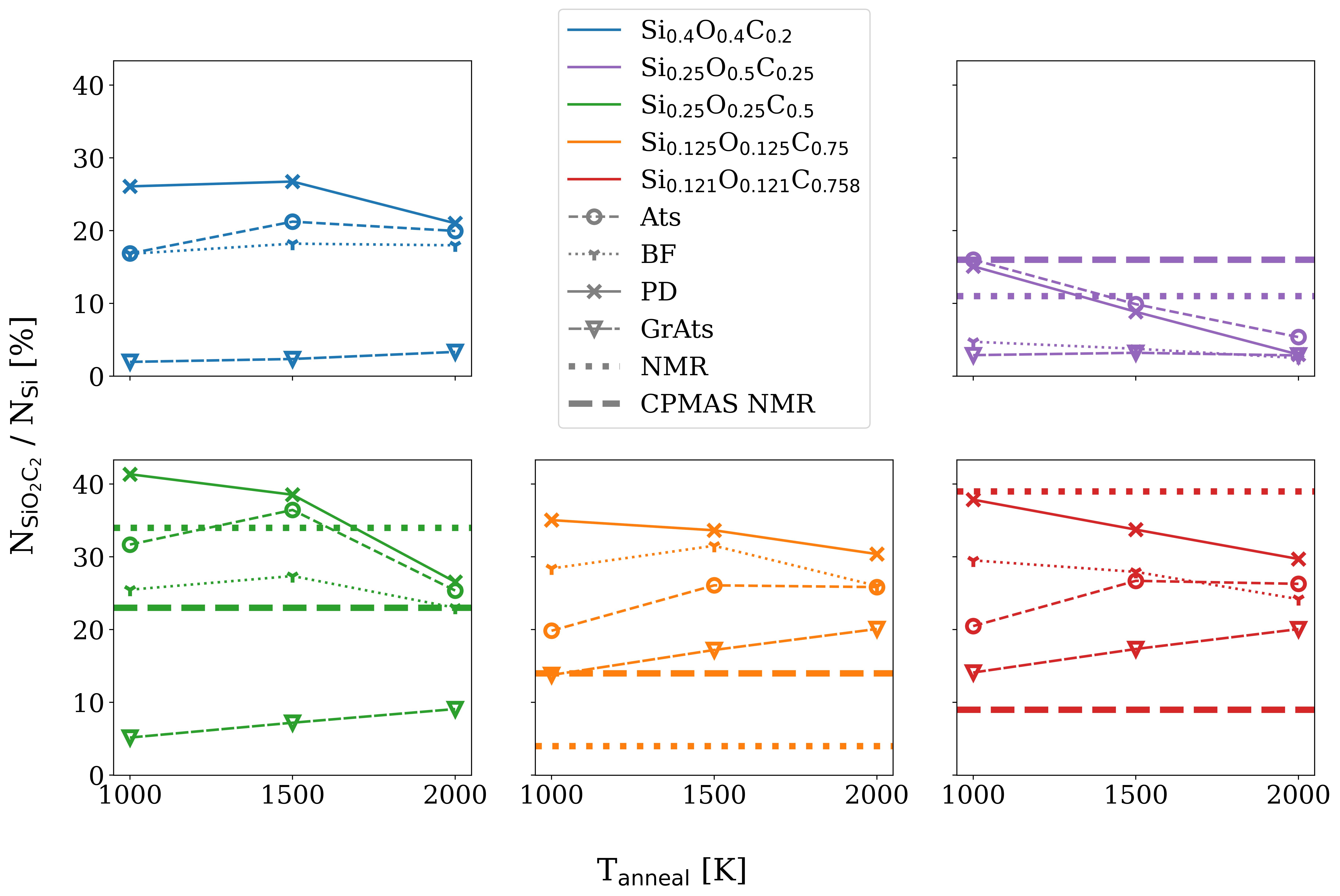}
    \caption{See text for more information}
    \label{fig:SiO2C2Frac}
\end{figure*}
\begin{figure*}[htbp!]
    \centering
    \includegraphics[width=0.9\linewidth]{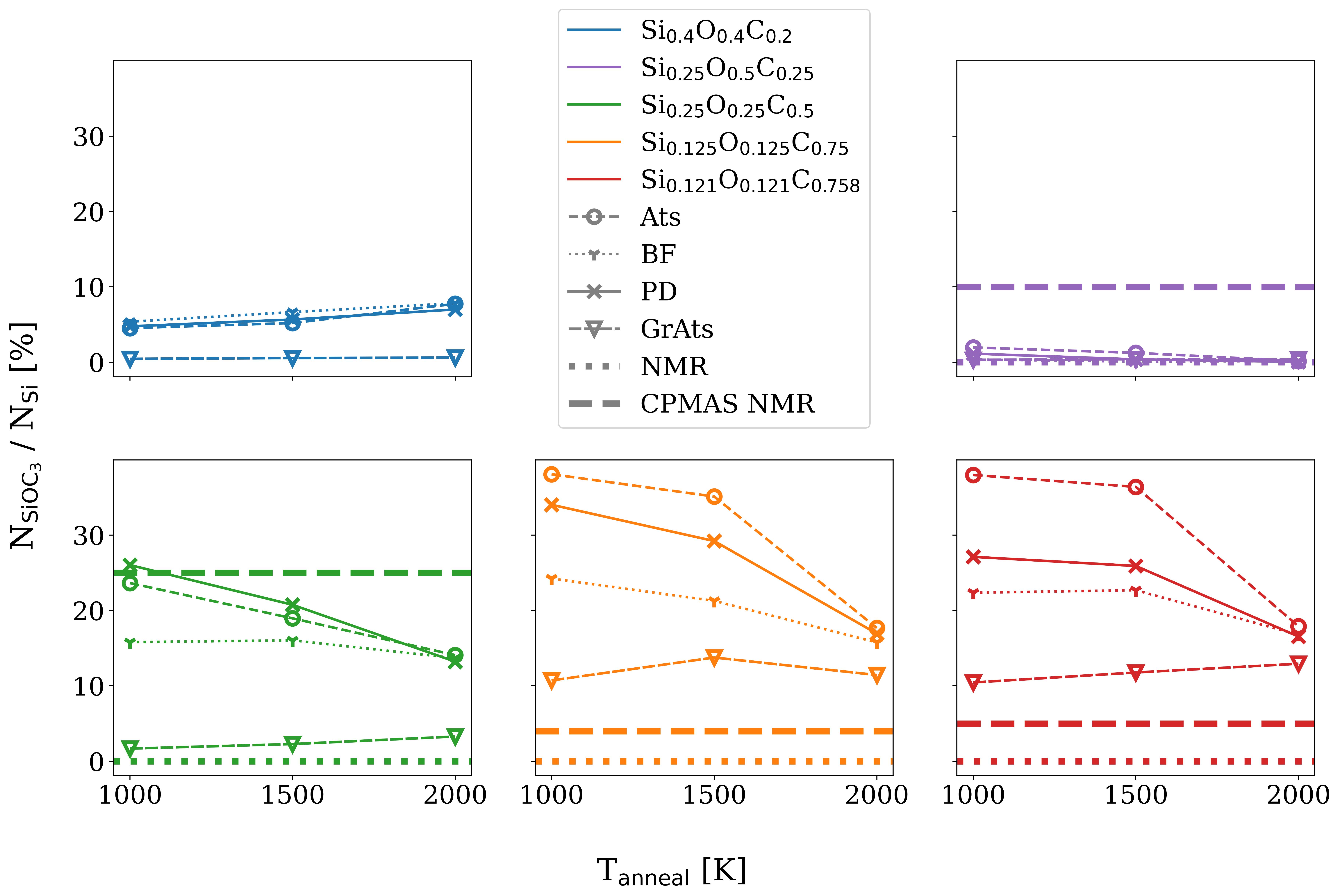}
    \caption{See text for more information}
    \label{fig:SiOC3Frac}
\end{figure*}
\begin{figure*}[htbp!]
    \centering
    \includegraphics[width=0.9\linewidth]{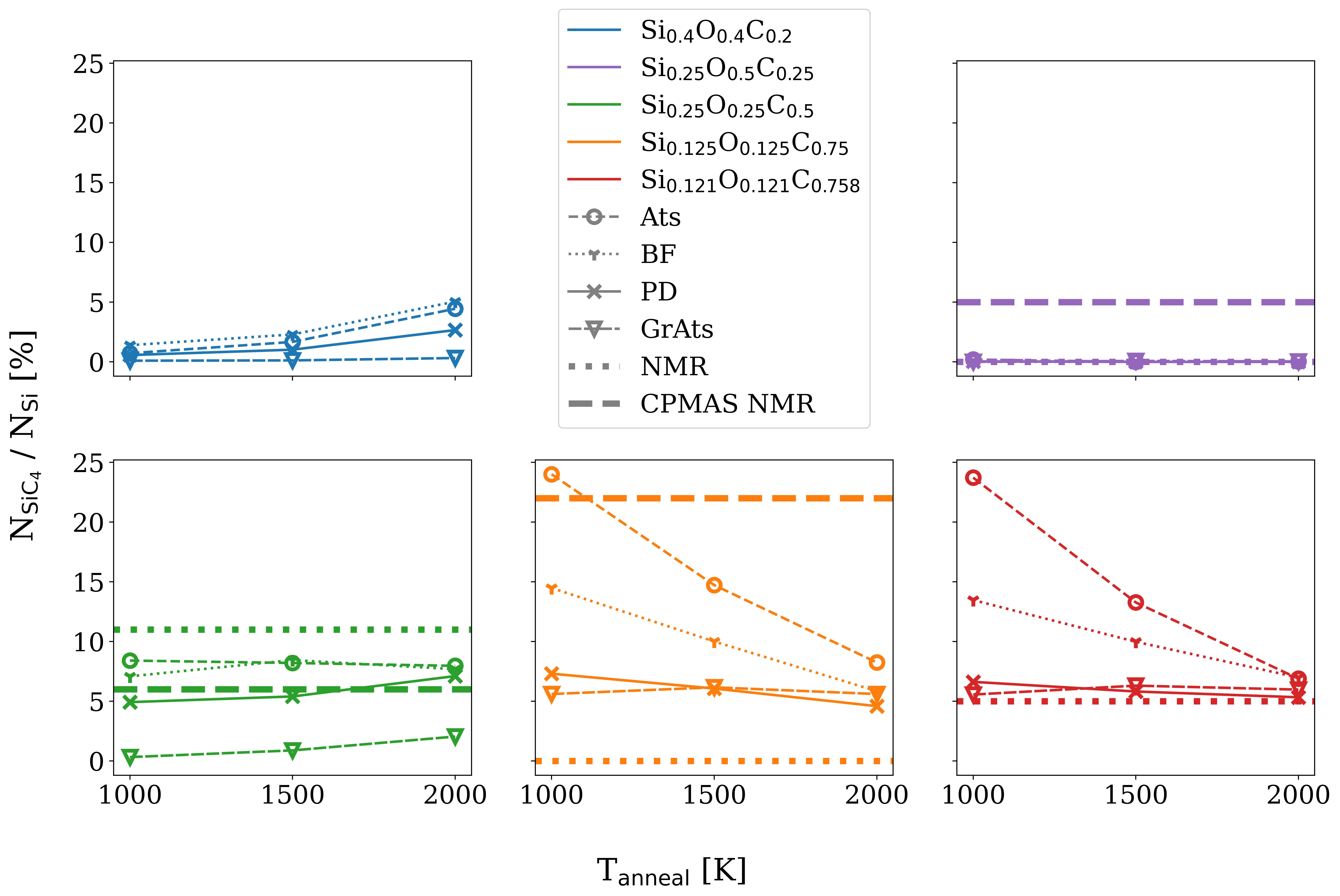}
    \caption{See text for more information}
    \label{fig:SiC4Frac}
\end{figure*}

Fractions of different SiO$_x$C$_{4-x}$ tetrahedra for the samples used in this work are shown in Fig. \ref{fig:SiO4Frac}-\ref{fig:SiC4Frac}.
Generally values differ significantly depending on the temperature and building blocks used to generate the structures.

The plots also include experimental values measured with $^{29}$Si CPMAS NMR (dashed horizontal lines) and Hahn-echo $^{29}$Si MAS NMR (dotted horizontal lines),
taken from \cite{sicSiCOCeramicsStorage}.
Please note that the experimental samples have significantly different compositions than the simulated ones,
because several gaseous products can evaporate during the polymer based synthesis route.
The SiO$_4$ and SiO$_3$C fractions are typically slightly underestimated.
For SiO$_2$C$_2$ and SiC$_4$ the agreement is better, while SiO$C_3$ is overestimated especially for the carbon rich structures with SILRES604 and RD684
composition.
A major reason for these differences is presumably the already mentioned difference in compositions
in combination with the short time scales achievable in MD simulations.

\section{Young's modulus from elastic tensor \label{sec:YoungsModulus}}

\begin{figure*}[htbp!]
    \centering
    \includegraphics[width=0.8\linewidth]{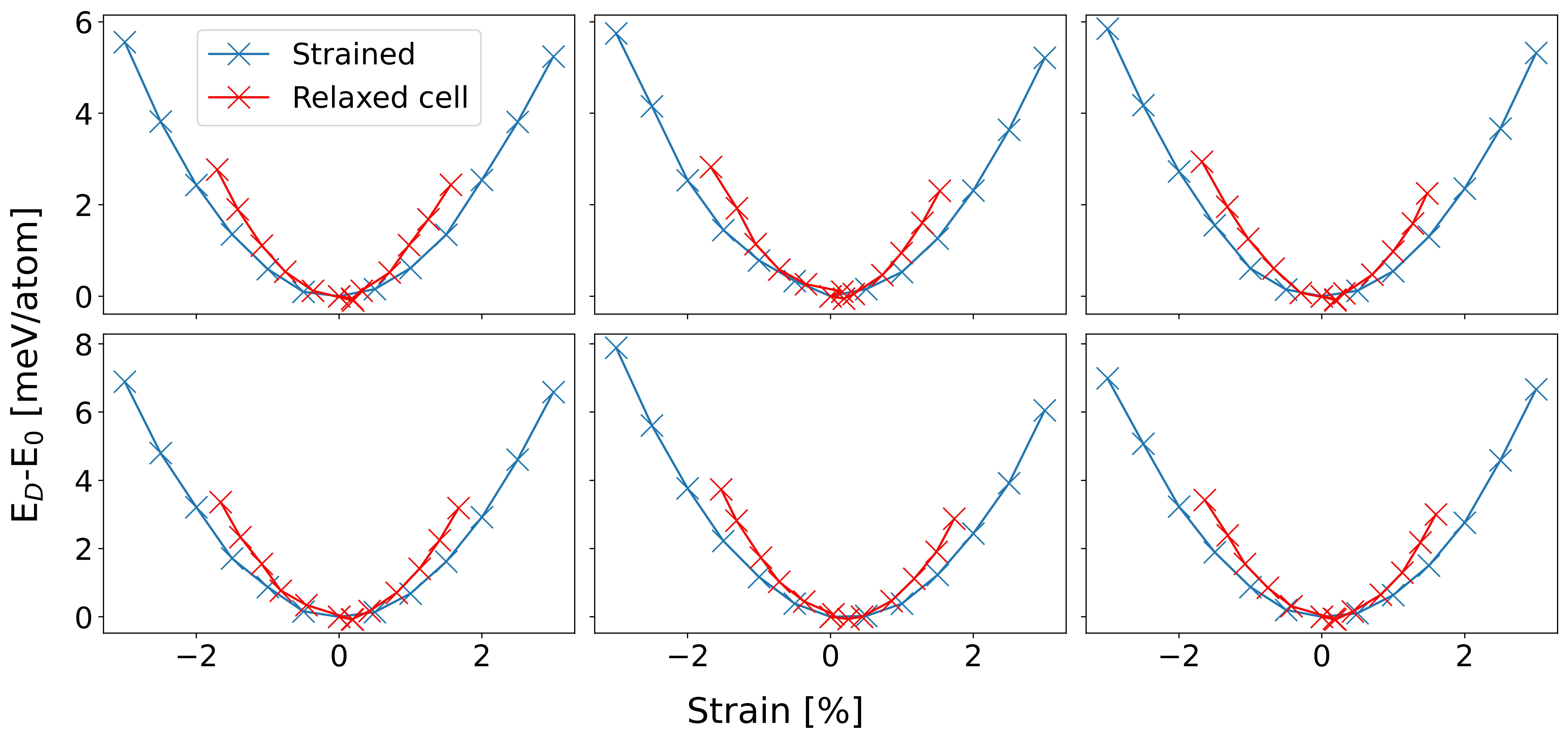}
    \caption{Strain-energy relation of Si$_2$O$_2$C structure based on single atoms.}
    \label{fig:verifyStoichometricFromAtoms}
\end{figure*}

\begin{figure*}[htbp!]
    \centering
    \includegraphics[width=0.8\linewidth]{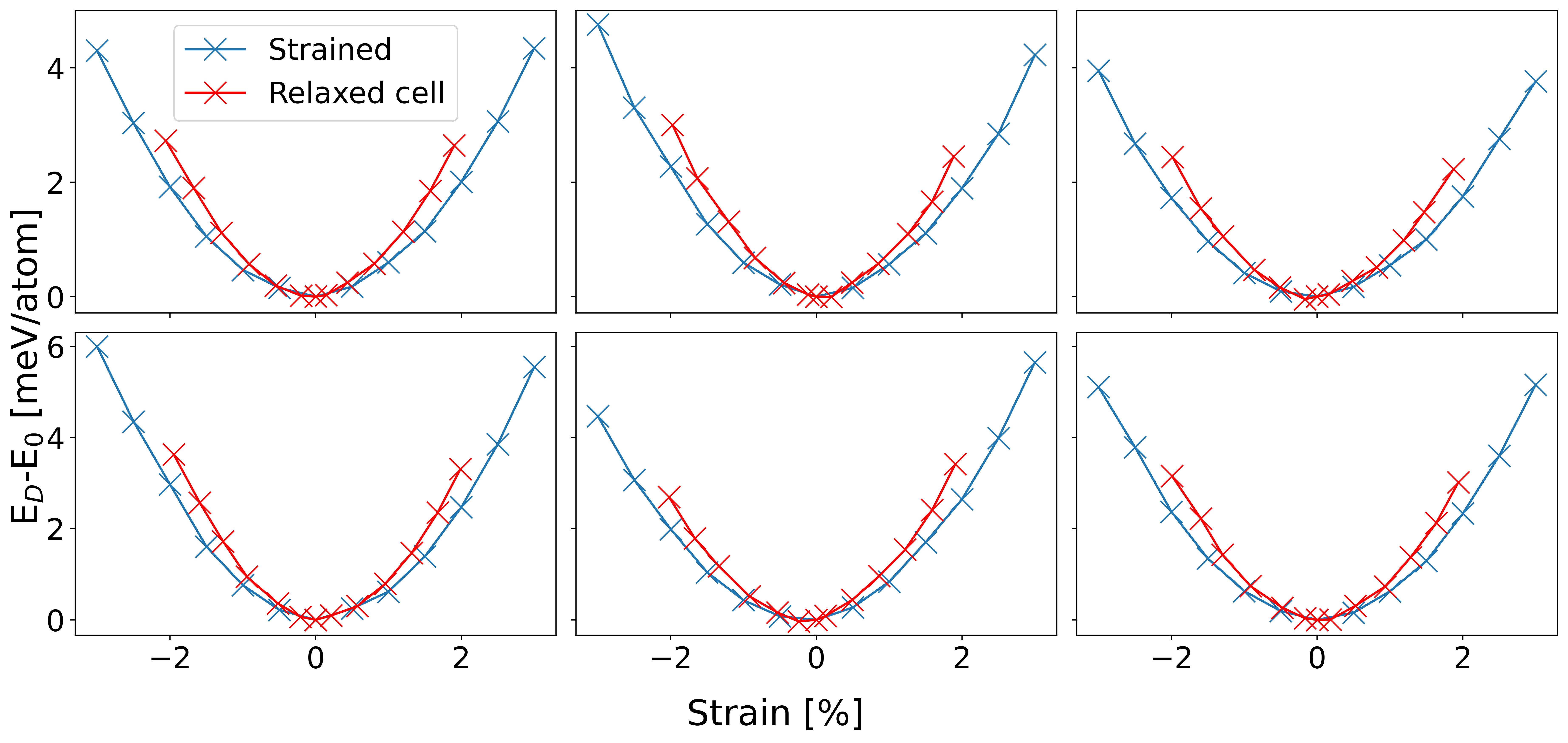}
    \caption{Strain-energy relation of PMSQ structure based on graphite and atoms.}
    \label{fig:verifyPMSQFromGraphAts}
\end{figure*}

Figures \ref{fig:verifyStoichometricFromAtoms} to \ref{fig:verifyPMSQFromGraphAts} show the strain-energy relation of the 6 applied strain states for a subset of structures
and expected approximately harmonic behavior.
The elastic tensors are calculated by the linear strain-stress relation.

\begin{figure*}[htbp!]
    \centering
    \includegraphics[width=0.8\linewidth]{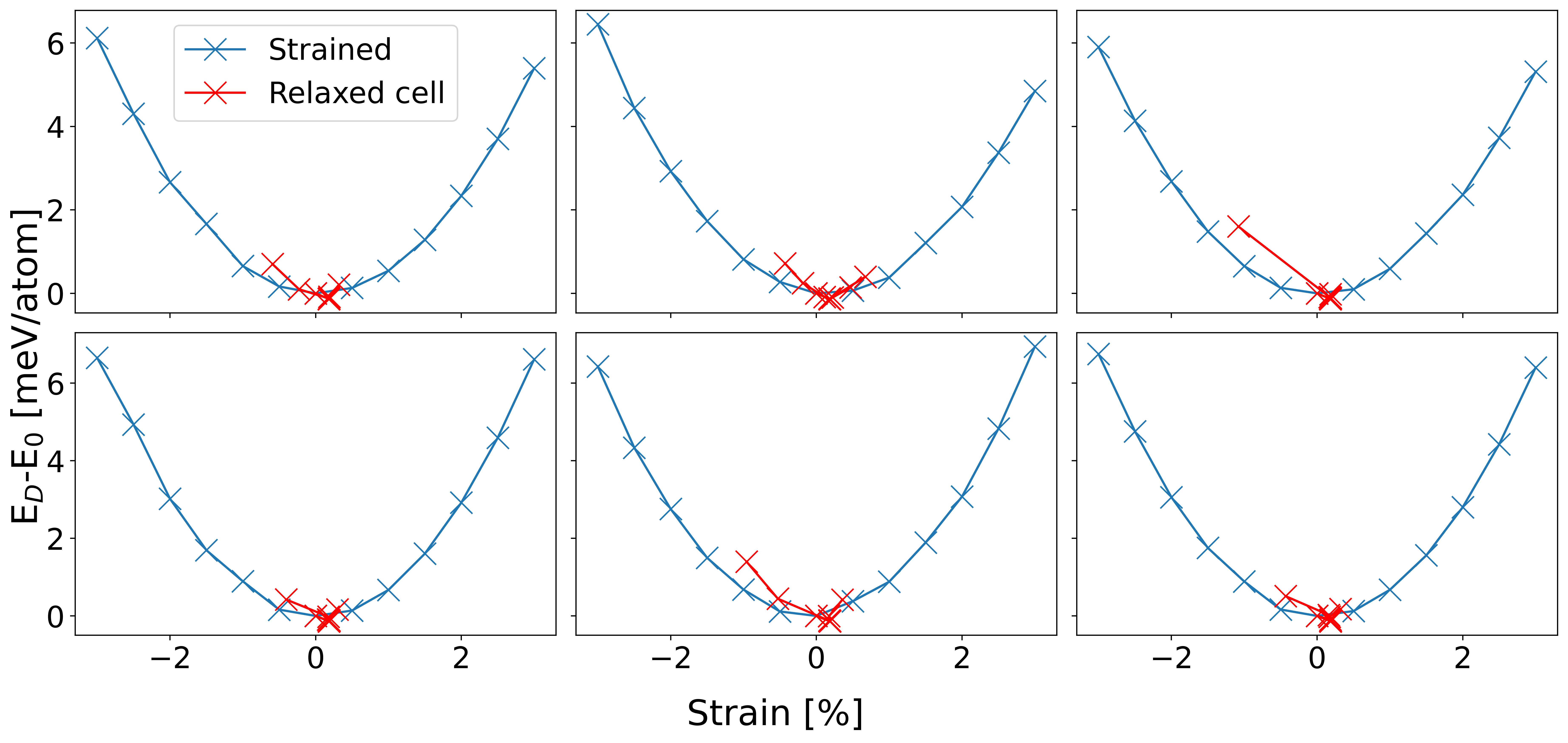}
    \caption{Strain-energy relation of RD-212 structure based on H-stripped polymers.}
    \label{fig:verifyRD212FromPoly}
\end{figure*}
\begin{figure*}[htbp!]
    \centering
    \includegraphics[width=0.8\linewidth]{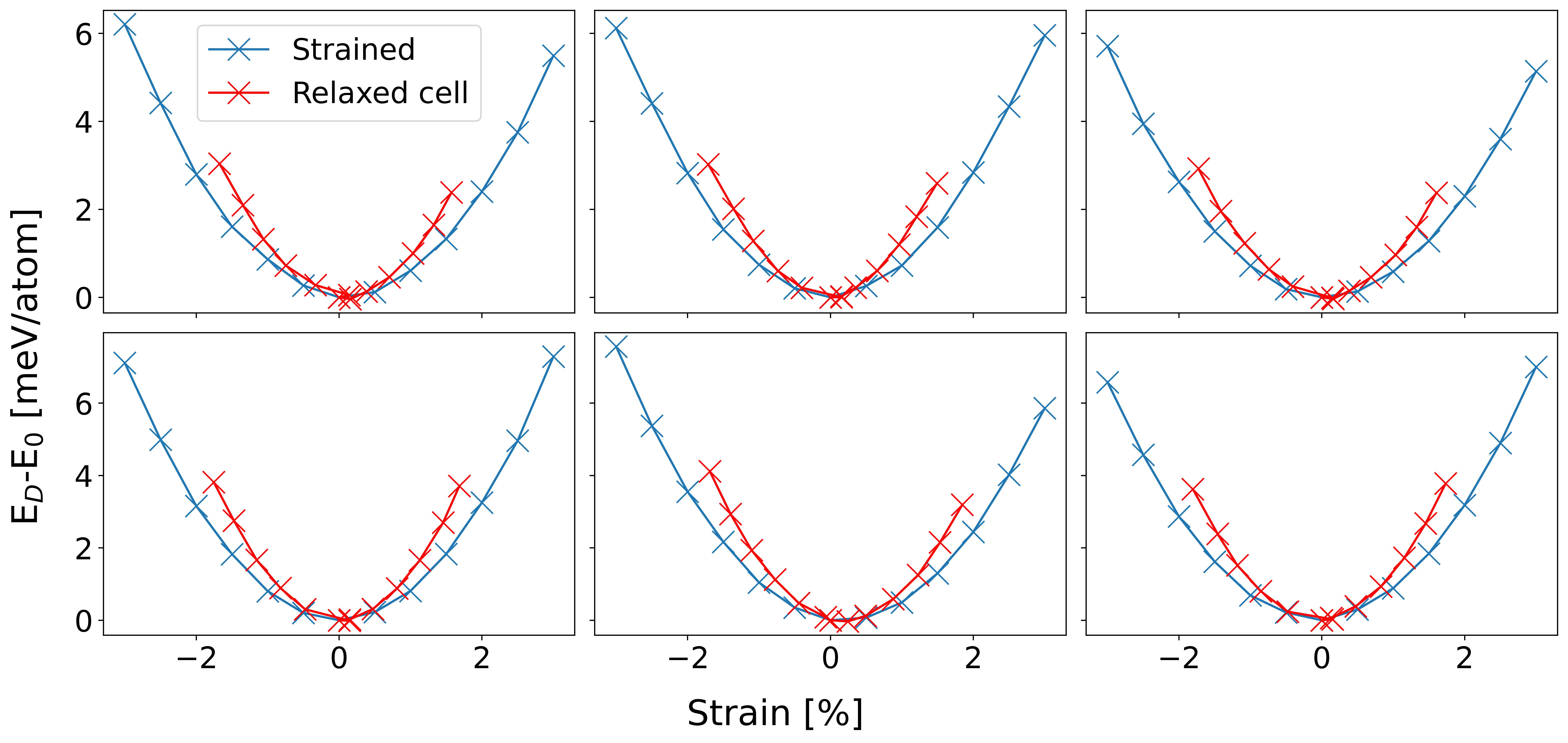}
    \caption{Strain-energy relation of SILRES-604 structure based on molecules.}
    \label{fig:verifySILRES604FromMols}
\end{figure*}

\section{Speed of ReaxFF and ACE}

\begin{figure}[tbp!]
    \centering
    \includegraphics[width=0.5\linewidth]{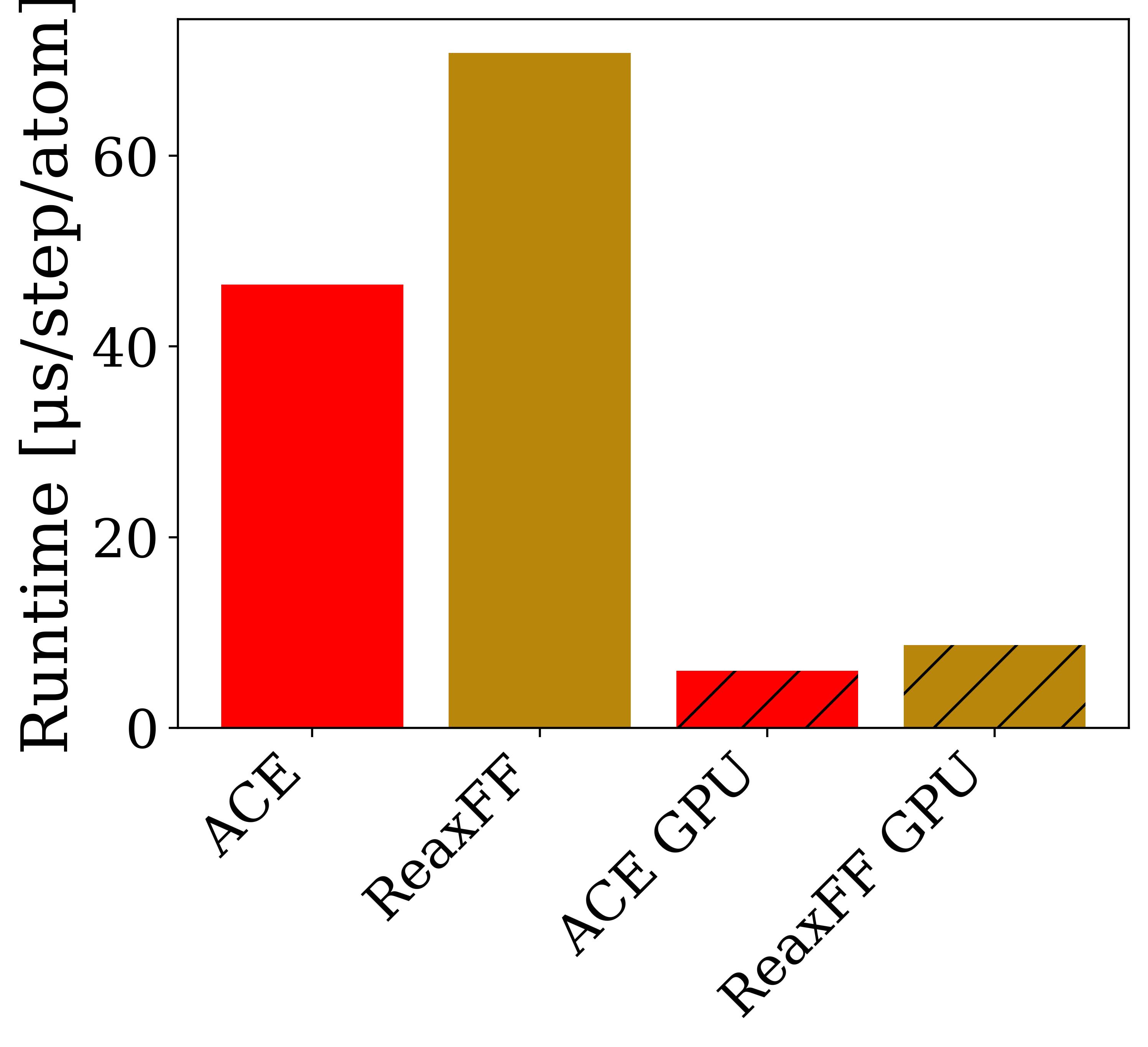}
    \caption{The computational costs of \ac{ACE} and ReaxFF potentials are about the same order of magnitude. See text for details.}
    \label{fig:ACEvsReaxFFSpeed}
\end{figure}

The computational cost of the fitted ACE potential and an exemplary ReaxFF potential \cite{newsomeOxidationSiliconCarbide2012} is about the same order of magnitude as shown in
Fig. \ref{fig:ACEvsReaxFFSpeed}.
For the evaluation of the runtime the LAMMPS code was employed to simulate an SiOC structure with 80896 atoms
on a single core of AMD Ryzen 5800X processor and an NVIDIA RTX 3060 GPU respectively.

In this example the fitted ACE potential is slightly faster than the ReaxFF.
However, we want to note that the runtime of ACE potentials massively depends
on the amount of basis functions used in the fit.
Similarly, the runtime of the ReaxFF potential depends on the frequency
with which charges are equilibrated and the employed cutoff,
so other tests could show both potentials
significantly faster or slower.

